\documentclass[fleqn,usenatbib]{mnras}

\usepackage[T1]{fontenc}
\usepackage{ae,aecompl}

\usepackage{eso-pic}

\AddToShipoutPictureBG*{%
  \AtPageUpperLeft{%
    \hspace{0.75\paperwidth}%
    \raisebox{-4.5\baselineskip}{%
}}}%

\usepackage{graphicx}	
\usepackage{amsmath}	
\usepackage{amssymb}	
\usepackage[shortlabels]{enumitem}

\usepackage{multicol}

\usepackage{newtxtext,newtxmath}

\title[MaNGA Photometric and Morphological Catalogs]{SDSS-IV DR17: Final Release of MaNGA PyMorph Photometric and Deep Learning Morphological Catalogs}

\author[ Dom\'inguez S\'anchez et al.]{
\parbox{\textwidth}{
\Large
H.~Dom\'{i}nguez S{\'a}nchez$^{1, 2}$\thanks{Corresponding author: \texttt{\rm \texttt{dominguez@ice.csic.es}}},
  B.~Margalef$^{3}$,
  M.~Bernardi$^{3}$, 
  M.~Huertas-Company$^{4, 5, 6, 7}$
  }
  \vspace{0.4cm}\\~\\
$^{1}$ Institute of Space Sciences (ICE, CSIC), Campus UAB, Carrer de Can Magrans, s/n, 08193 Barcelona, Spain\\
$^{2}$ Institut d’Estudis Espacials de Catalunya (IEEC), Carrer Gran Capità, 08034 Barcelona, Spain \\
$^{3}$ Department of Physics and Astronomy, University of Pennsylvania, Philadelphia, PA 19104, USA\\
$^{4}$ LERMA, Observatoire de Paris, PSL Research University, CNRS, Sorbonne Universit\'es, UPMC Univ. Paris 06,
F-75014 Paris, France\\
$^{5}$ University of Paris Denis Diderot, University of Paris Sorbonne Cit\'e (PSC), 75205 Paris Cedex 13, France\\
$^{6}$
Instituto de Astrof\'isica de Canarias, E-38200 La Laguna, Tenerife, Spain\\
$^{7}$
Departamento de Astrof\'isica, Universidad de La Laguna, E-38206 La Laguna, Tenerife, Spain\\
\vspace{-1cm}
}

\pubyear{2021}

\begin{document}
\label{firstpage}
\pagerange{\pageref{firstpage}--\pageref{lastpage}}
\maketitle

\begin{abstract}
  We present the MaNGA PyMorph photometric Value Added Catalogue (MPP-VAC-DR17) and the MaNGA Deep Learning Morphological  VAC (MDLM-VAC-DR17) for the final data release of the MaNGA survey, which is part of the SDSS Data Release 17 (DR17).
  The MPP-VAC-DR17 provides photometric parameters from S{\'e}rsic and S{\'e}rsic+Exponential fits to the 2D surface brightness profiles of the MaNGA DR17 galaxy sample in the $g$, $r$, and $i$ bands (e.g. total fluxes, half light radii, bulge-disk fractions, ellipticities, position angles, etc.).
  The MDLM-VAC-DR17 provides Deep Learning-based morphological classifications for the same galaxies. The MDLM-VAC-DR17 includes a number of morphological properties: e.g., a T-Type,  a finer separation between elliptical and S0, as well as the identification of edge-on and barred galaxies. While the MPP-VAC-DR17 simply extends the MaNGA PyMorph photometric VAC published in the SDSS Data Release 15 (MPP-VAC-DR15) to now include galaxies which were added to make the final DR17, the MDLM-VAC-DR17 implements some changes and improvements compared to the previous release (MDLM-VAC-DR15): namely, the low-end of the T-Types are better recovered in this new version. The catalogue also includes  a separation between  Early- or Late-type (ETG, LTG), which classifies the two populations in a complementary way to the T-Type, especially at the intermediate types (-1 < T-Type < 2), where the T-Type values show a large scatter. In addition, $k-$fold based uncertainties on the classifications are also provided. To ensure robustness and reliability, we have also visually inspected all the images. We describe the content of the catalogues and show some interesting ways in which they can be combined.

\end{abstract}




\section{Introduction}

As we enter the age of large galaxy samples for which spatially resolved spectroscopic information is available  thanks to Integral Field Spectroscopic surveys like  ATLAS$^{\rm 3D}$ \citep{Cappellari2011}, CALIFA \citep{Sanchez2012}, or SAMI \citep{Allen2015}, it is useful to have accompanying analyses of the associated photometry.  
\cite{Fischer2019} describe a step in this direction: they provide imaging-based morphological information, as well as one- and two-component fits to the two-dimensional surface brightness distributions of the galaxies in an early release  (SDSS-DR15, \citealt{Aguado2019}) of the MaNGA (Mapping Nearby Galaxies at Apache Point Observatory; \citealt{Bundy2015}) Survey.  Now that the survey is complete, the main goal of the present work is to extend that analysis to all the $\sim 10^4$ nearby ($z\sim 0.03$) galaxies in it.  This has culminated in the production of two `value-added' catalogs (VACs) which are part of the SDSS-DR17 release (SDSS collaboration, in prep.):
the MaNGA PyMorph photometric Value Added Catalogue  (hereafter MPP-VAC-DR17) and the MaNGA Deep Learning Morphology Value Added catalog (hereafter MDLM-VAC-DR17) which summarize the photometric and deep-learning based morphological information for the MaNGA galaxies.

MaNGA is a component of the Sloan Digital Sky Survey IV (\citealt{Blanton2017}; hereafter SDSS~IV).  \cite{Wake2017} describe how the MaNGA galaxies were selected from the SDSS footprint.  Integral Field Unit (IFU) technology allows the MaNGA survey to obtain detailed kinematic and chemical composition maps of these galaxies (e.g. \citealt{Gunn2006, Drory2015, Law2015, Law2016, Smee2013, Yan12016,Yan22016, Greene2017, Graham2018}).

For reasons discussed in \cite{Fischer2017}, we do not use the SDSS pipeline photometry of these objects.  Rather, we use the significantly more accurate PyMorph analysis described in a series of papers (\citealt{Vikram2010, Meert2013, Meert2015, Meert2016, Bernardi2014}).  PyMorph provides one and two-component fits to the two-dimensional surface brightness distributions of MaNGA galaxies, and was used to produce the MPP-VAC of the galaxies in DR15 \citep{Fischer2019}.  The MPP-VAC-DR17 which we describe below extends this to include all the objects in the completed MaNGA survey.

We also provide the MDLM-VAC-DR17, which includes Deep Learning-based morphological classifications (the methodology is described in detail by \citealt{DS2018}) for the same galaxies.  In contrast to the photometric MPP-VACs, in which the main difference between the DR15 and DR17 versions is sample size, the MDLM-VAC-DR17 includes some improvements in methodology and content with respect to DR15, which we describe below.  

Note that the MaNGA data was only used for the identification of the sources included in the two VACs presented in this paper. Both the two-dimensional fits to the surface brightness distributions and the morphological classifications are based on the SDSS imaging data (DR15 for the MPP-VAC and DR7 for the MDLM-VAC).

Section~\ref{sec:mpp} describes the minor changes we have made when reporting the photometric parameters listed in the MPP-VAC:  see \cite{Fischer2019} for a detailed discussion of how these PyMorph-based parameters were determined, and how they compare with previous work. Section~\ref{sec:mdlm} describes our morphological classification scheme and the MDLM-VAC-DR17 which results. Section~\ref{sec:photmorph} combines our MPP- and MDLM-VACs to show how the photometric parameters correlate with morphology. A final section summarizes.

\section{MaNGA PyMorph Photometric Value Added Catalog (MPP-VAC-DR17)}\label{sec:mpp}

The MPP-VAC-DR17\footnote{www.sdss.org/dr17/data\textunderscore access/value-added-catalogs/?vac\textunderscore id=manga-pymorph-dr17-photometric-catalog} is one of the value added catalogs available online of the completed MaNGA survey, which is part of the SDSS-DR17 release\footnote{www.sdss.org/dr17/data\textunderscore access/value-added-catalogs/}. It is similar to the MPP-VAC-DR15 \citet[F19 hearafter]{Fischer2019} published as part of the SDSS-DR15 release \citep{Aguado2019}. The MPP-VAC-DR17 is updated to include all the galaxies in the final MaNGA release. Some PLATE-IFU entries are re-observations of the same galaxy so the catalog also provides three variables which identify galaxies with multiple MaNGA spectroscopic observations (see DUPL-GR, DUPL-N, and DUPL-ID).  Although the number of entries is 10293, the actual number of different galaxies in this catalogue is 10127. The structural parameters and morphological classifications included in the VACs are identical for the duplicate observations. 

The MPP-VAC-DR17  also includes one minor technical change regarding how the position angle of each image is reported. The position angle PA (from PyMorph) given in this catalogue is with respect to the camera columns in the SDSS ``fpC'' images (which are not aligned with the North direction); to convert to the usual convention where North is up, East is left \footnote{note that the MaNGA datacubes have North up East right}, set PA(MaNGA) = (90 - PA) - SPA, where SPA is the SDSS camera column position angle with respect to North reported in the primary header of the ``fpC'' SDSS images. PA (MaNGA) is defined to increase from East towards North. In contrast to the MPP-VAC-DR15 release where the SPA angles were provided in a separate file, the MPP-VAC-DR17 catalog includes the SPA angles.

Except for this change, MPP-VAC-DR17 is similar in format to MPP-VAC-DR15.  In particular, Table~1 in F19 describes the content of the catalog, which is in the FITS file format and includes 3 HDUs. Each HDU lists the parameters measured in the $g$, $r$, and $i$ bands, respectively.  These include the luminosity, half-light radius, S{\'e}rsic index, etc. for single S{\'e}rsic (Ser) and two-component S{\'e}rsic+Exponential (SerExp) profiles -- from fitting the 2D surface brightness profiles of each galaxy. Although for most galaxies the Exponential component is a disk, for the most luminous galaxies, it represents a second component which needs not be a disk.

None of the algorithms has changed since DR15, so the discussion in F19 about how photometric parameters were determined, remains appropriate.  In particular, we still use the fitting algorithm called PyMorph (\citealt{Vikram2010, Meert2013, Meert2015, Meert2016, Bernardi2014}), a Python based code that uses Source Extractor (SExtractor; \citealt{BA1996}) and GALFIT \citep{Peng2002} to estimate the structural parameters of galaxies.  Likewise, decisions about refitting (Section 2.1.1 in F19), when to `flip' the two components of a SerExp fit (Section 2.1.3 in F19), and how to truncate the profiles (Section 2.1.4 in F19) are all the same as before, as is the (visual inspection-based) flagging system, that indicates which fit is to be preferred for scientific analyses (see discussion in Section 2.2 of F19).  We urge users to pay attention to the preferences expressed by FLAG$\_$FIT:  FLAG$\_$FIT$=$1 means that the single-S{\'e}rsic fit is preferred (the SerExp fit may be unreliable), FLAG$\_$FIT$=$2 means that the SerExp fit is preferred (the Ser fit may be unreliable), FLAG$\_$FIT$=$0 means that both Ser and SerExp fits are acceptable, and FLAG$\_$FIT$=$3 means that none of the fits were reliable and so no parameters are provided. Table~\ref{tabCat} lists the fraction of objects for each FLAG$\_$FIT type in the SDSS $g$, $r$, and $i$ bands.

\begin{table}
\centering
FRACTION OF GALAXIES\\
\begin{tabular}{cccc}
  \hline
 Band & Sersic fit failed & SerExp fit failed  &  Both fits failed \\
 & (FLAG$\_$FAILED$\_$S = 1) & (FLAG$\_$FAILED$\_$SE = 1) & (FLAG$\_$FIT = 3) \\
\hline
       $g$  &  0.069  &   0.065 &   0.038 \\
       $r$  &  0.065  &  0.058  &  0.034  \\
       $i$  &  0.077 &   0.062  &  0.037 \\
 \hline
 \hline
 \end{tabular}
 Galaxies with successful fits (FLAG$\_$FIT $\ne$ 3) better described by\\
 \begin{tabular}{cccc}
 \hline
Band & 1-component & 2-components & Both \\
 & (FLAG$\_$FIT = 0) & (FLAG$\_$FIT = 1) & (FLAG$\_$FIT = 2) \\
 \hline
    $g$ &   0.103   &   0.586    &  0.312 \\
    $r$ &   0.106   &   0.567    &  0.327 \\
    $i$ &   0.104   &   0.569    &  0.327 \\
 \hline
 \hline
\end{tabular}
\caption{Top: Fraction of galaxies which do not have PyMorph parameters from S{\'e}rsic (FLAG$\_$FAILED$\_$S = 1), SerExp (FLAG$\_$FAILED$\_$SE = 1) or both (FLAG$\_$FIT = 3) in the SDSS $g$, $r$ and $i$ bands. Bottom: Fraction of galaxies which are better described by 1 component S{\'e}rsic fit (FLAG$\_$FIT = 1), 2 components SerExp fit (FLAG$\_$FIT = 2) or for which both fits are equally acceptable (FLAG$\_$FIT = 0). 
}
\label{tabCat}
\end{table}

\begin{figure}
  \centering
  \includegraphics[width=0.9\linewidth]{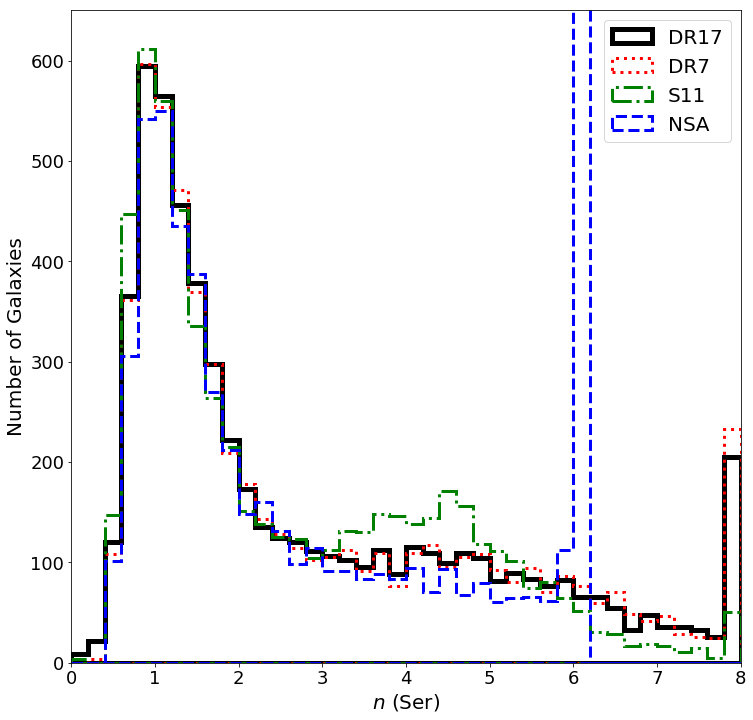}
  \caption{Distribution of $n$ from single S{\'e}rsic fits to the $r$-band surface brightness profiles of the objects in our sample (DR17),  compared to the corresponding distribution from \citet[DR7]{Meert2015},  \citet[S11]{Simard2011} and the NASA-Sloan Atlas catalog (NSA; {\tt nsatlas.org}). Our analysis limits $n \le 8$, whereas the S11 analysis allows $0.5 \le n \le 8$, and NSA does not allow $n > 6$. This explains the spike at $n = 6$ where NSA has $1709$ galaxies.}
  \label{fig:nSer}
\end{figure}

\begin{figure}
  \centering
  \includegraphics[width=0.9\linewidth]{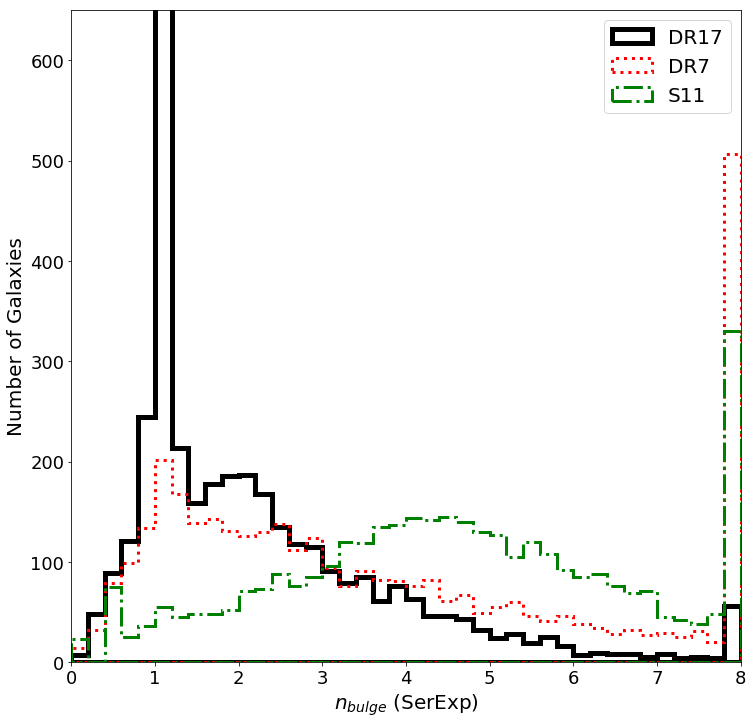}
  \caption{Same as previous figure, but for $n_{\rm bulge}$ of the two-component SerExp fits. See \citet{Fischer2019} for discussion of the obvious differences with respect to S11. Similarly to DR15, our DR17 analysis has several more galaxies with $n_{\rm bulge}=1$ but many fewer $n_{\rm bulge}=8$ compared to the DR7 analysis, as a result of our eye-ball motivated refitting and flipping. The spike $n_{\rm bulge}=1$ for DR17 extends to 735 galaxies.}
 \label{fig:nSerExp}
\end{figure}

The flags FLAG$\_$FAILED$\_$S $=1$ or FLAG$\_$FAILED$\_$SE $=1$ indicate failed S{\'e}rsic or SerExp fit, respectively. Failures can happen for several reasons: contamination, peculiarity, bad-image, or bad model fit. The numbers in the top half of Table~\ref{tabCat} give the fraction of objects without photometric measurements for the different bands. About 7\% of the objects do not have parameters from the S{\'e}rsic and SerExp fits, respectively. About 4\% of these objects do not have any PyMorph photometric parameters (i.e. FLAG$\_$FIT $=3$).

Figures~\ref{fig:nSer} and~\ref{fig:nSerExp} show the distributions of the S{\'e}rsic index $n$ and $n_{\rm bulge}$ in our single and two-component fits.  These are very similar to Figures~12 and~14 in F19, illustrating that other than the factor of 2 increase in sample size (from DR15 to DR17) the trends are unchanged.  In particular, our reductions do not show a preference for $n=6$ (in contrast to NSA which does not allow $n>6$), or for $n=4$ or $n_{\rm bulge}=4$ (in contrast to S11). Similarly to DR15, our DR17 analysis has several more galaxies with $n_{\rm bulge}=1$ but many fewer $n_{\rm bulge}=8$ compared to the DR7 analysis, as a result of our eye-ball motived refitting and flipping. Likewise, we have repeated all the other tests and comparisons shown in F19, but now for the full DR17 sample, finding consistent results with our DR15 analysis, so we do not show them here.

\section{MaNGA Deep Learning Morphology Value Added catalog (MDLM-VAC-DR17)}\label{sec:mdlm}

The MaNGA Deep Learning Morphology DR17 VAC (MDLM-VAC-DR17) provides morphological classifications for the final MaNGA galaxy sample  (which is part of the SDSS-DR17 release) using an automated classification based on supervised deep learning. It extends the `MaNGA  Deep Learning Morphology DR15 VAC', described in F19, to now include galaxies which were added to make the final DR17.  In addition, as we describe in the following sections, it incorporates some changes and improvements with respect to the DR15 version.

\begin{figure*}
\setlength{\columnsep}{20pt}
    \includegraphics[width=0.5\linewidth]{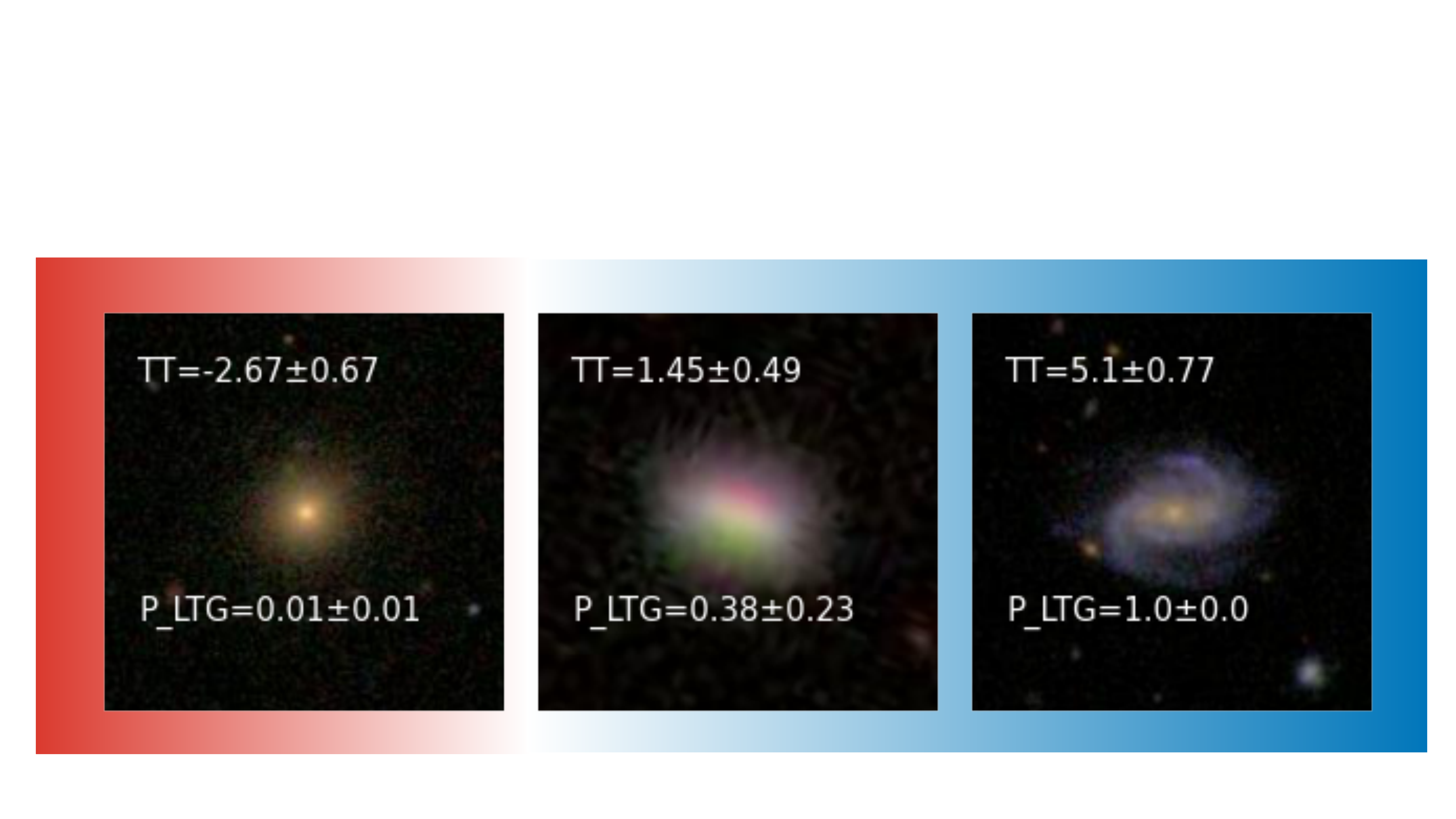}\par 
    \includegraphics[width=\linewidth]{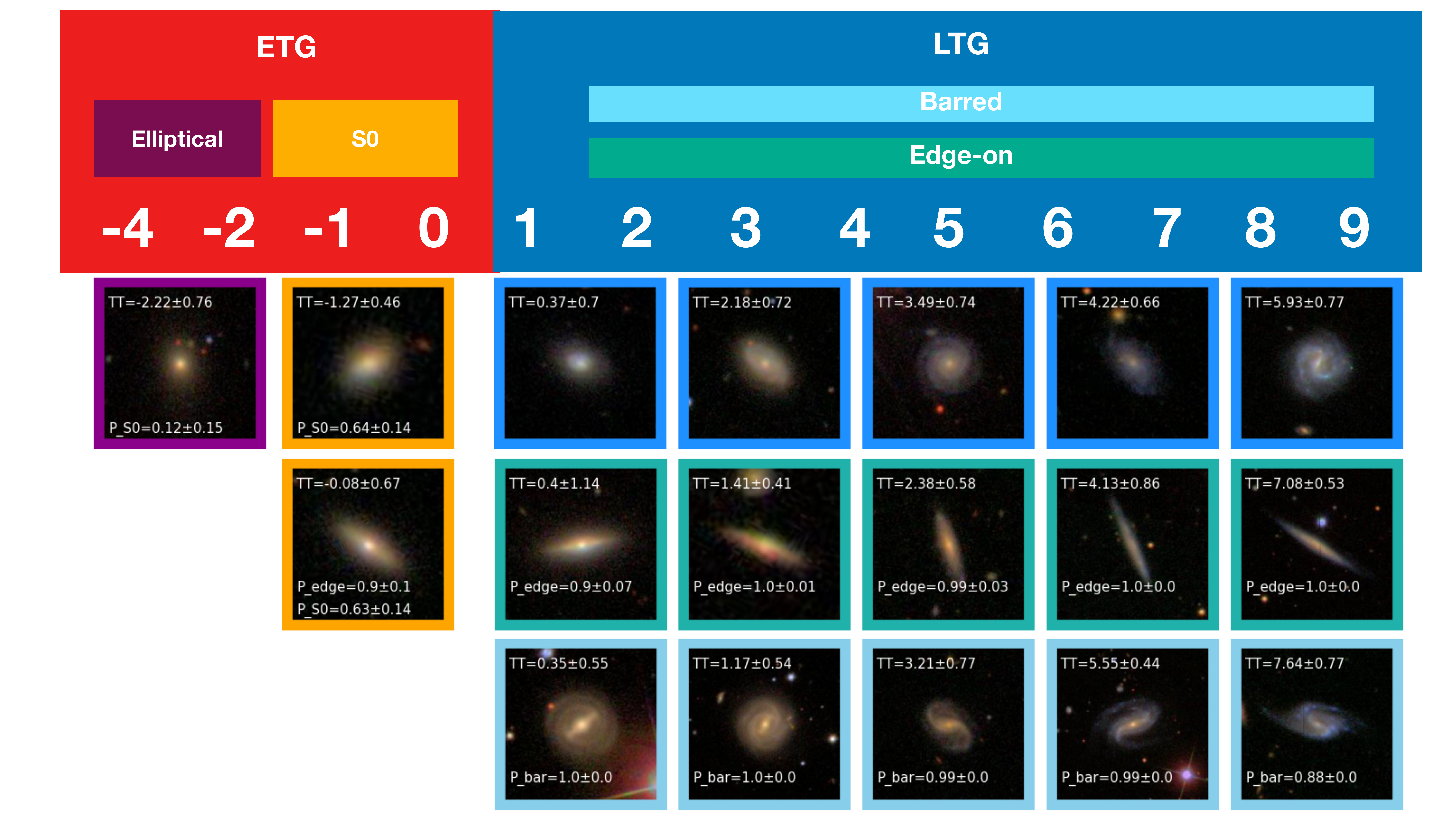}\par 
\caption{Schematic representation of the morphological classification presented in the MDLM-VAC.  It includes a T-Type ranging from -4 to $\sim$9, where the transition between early and late types happens around T-Type$\sim$0.  A complementary binary classification separates ETGs and LTGs (see discussion in the text for the differences between these two classifications). Three further binary classifications a) separate E from S0 -- this separation is only meaningful for galaxies with T-Type < 0, b) identify edge-on galaxies b) identify galaxies with bar features. The cutouts show examples of galaxies of different types, with classifications values shown in white and according to their frame colours. The cutouts are proportional to the size of each galaxy ($\sim$5$\times$R$_{90}$). The top most cutouts show a galaxy classified as ETG by both the T-Type and P$_{\rm LTG}$ models (left), a galaxy classified as LTG by both the T-Type and P$_{\rm LTG}$ models (right) and a galaxy with T-Type > 0 and P$_{\rm LTG}$ < 0.5 -- see discussion realated to these galaxies in section \ref{sect:P-LTG}.} \label{fig:TT1}
\end{figure*}

The morphological classifications were obtained following the methodology explained in detail in \citet[hereafter DS18]{DS2018}. Briefly, for each classification task, we trained a convolutional neural network (CNN) using as input the RGB cutouts downloaded from the SDSS-DR7 server\footnote{http://casjobs.sdss.org/ImgCutoutDR7/} with a variable size that is proportional to the Petrosian radius of the galaxy (5$\times$R$_{90}$\footnote{R$_{90}$ from NSA catalogue}). The cutouts are then re-sampled to 69$\times$69 pixels, which are the dimensions used to feed the CNN -- note that by doing this the pixel scale varies from one galaxy to another. The counts in each pixel are normalized by the maximum value in that cutout for that particular color band. As this value is different in each band, this step prevents color information from playing a role in the morphological classifications, and potentially biasing studies of color-morphology relations. We refer the reader to DS18 for further details.

\subsection{Classification scheme}

The classification scheme of the morphological catalogue is presented in Figure \ref{fig:TT1}. We provide a T-Type value, which ranges from -4 to 9, and was obtained by training the CNN in regression mode based on the T-Types from \citet[N10 herafter]{Nair2010} catalogue.  N10 presents visual classifications for 14034 galaxies from SDSS up to $m_{g} <$ 16 mag.  We only use galaxies with confident classifications for training [T-Type flag = 0, i.e. $\sim$96$\%$ of the sample].  While the N10 T-Type values are integers running from [-5, 10], none of their galaxies have T-Type values of -4, -2, or -1.  We reassigned T-Type values by shifting them and filling the gaps in our training labels, as doing so helps the model to converge.

In general, T-Type $<$ 0 corresponds to ETGs while T-Type $>$ 0 corresponds to LTGs.  Following F19, we sometimes subdivide LTGs into S1 ($0\le$ T-Type $\le 3$) and S2 (T-Type $>3$) -- see  section \ref{sect:P-LTG}. 

The catalogue provides two other binary classifications which were trained with  N10-based labels:\\
 - P$_{\rm LTG}$, which separates ETGs from LTGs\\
 - P$_{\rm S0}$, which separates pure ellipticals (E) from S0s\\

For the  P$_{\rm LTG}$ model, we labelled as positive examples those with T-Type $>$ 0  and as negative examples those with T-Type $\le 0$ (from N10). This classification complements the T-Type by providing a cleaner separation between ETGs and LTGs, specially at intermediate T-Types, where the scatter of the T-Type model is larger (see discussion in Section \ref{sect:T-Type}).  For the  P$_{\rm S0}$ model, we used as training sample only galaxies with T-Type $<$ 0 and labelled as positive examples those with -5 $<$ T-Type $<$ 0  and as negative examples those with T-Type $=$ -5 (i.e., pure E according to N10). 

The catalogue also provides two additional binary classifications:\\
 - P$_{\rm edge-on}$, which identifies edge-on galaxies\\
 - P$_{\rm bar}$, which identifies barred galaxies.\\
 
The value reported in the catalogue is the probability of being a positive example (edge-on or barred galaxy, respectively).  These are based on the Galaxy Zoo 2 \cite[][GZ2 hereafter]{Willett2013} labels.  GZ2 is a citizen science project with morphological classifications of 304122 galaxies drawn from SDSS up to m$_{r} < 17$. Following DS18, the training sample was composed of galaxies with robust classifications, i.e., at least 5 votes and weighted fraction values greater than 0.8 (for the `yes' or `no' answers in each  classification task). See DS18 for further details. 

\begin{figure*}
\setlength{\columnsep}{1pt}
\begin{multicols}{2}
    \includegraphics[width=0.8\linewidth]{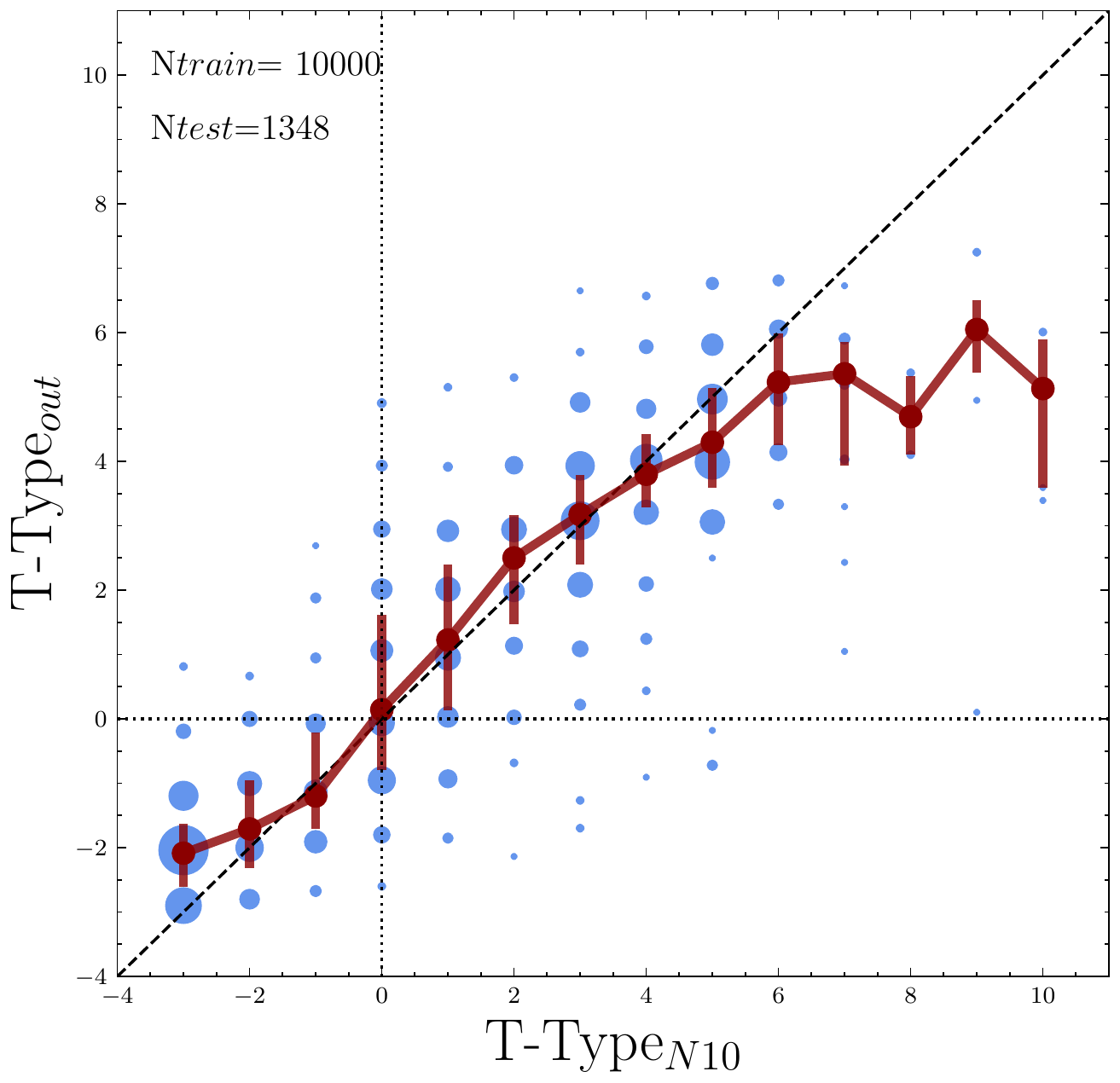}\par 
    \includegraphics[width=0.8\linewidth]{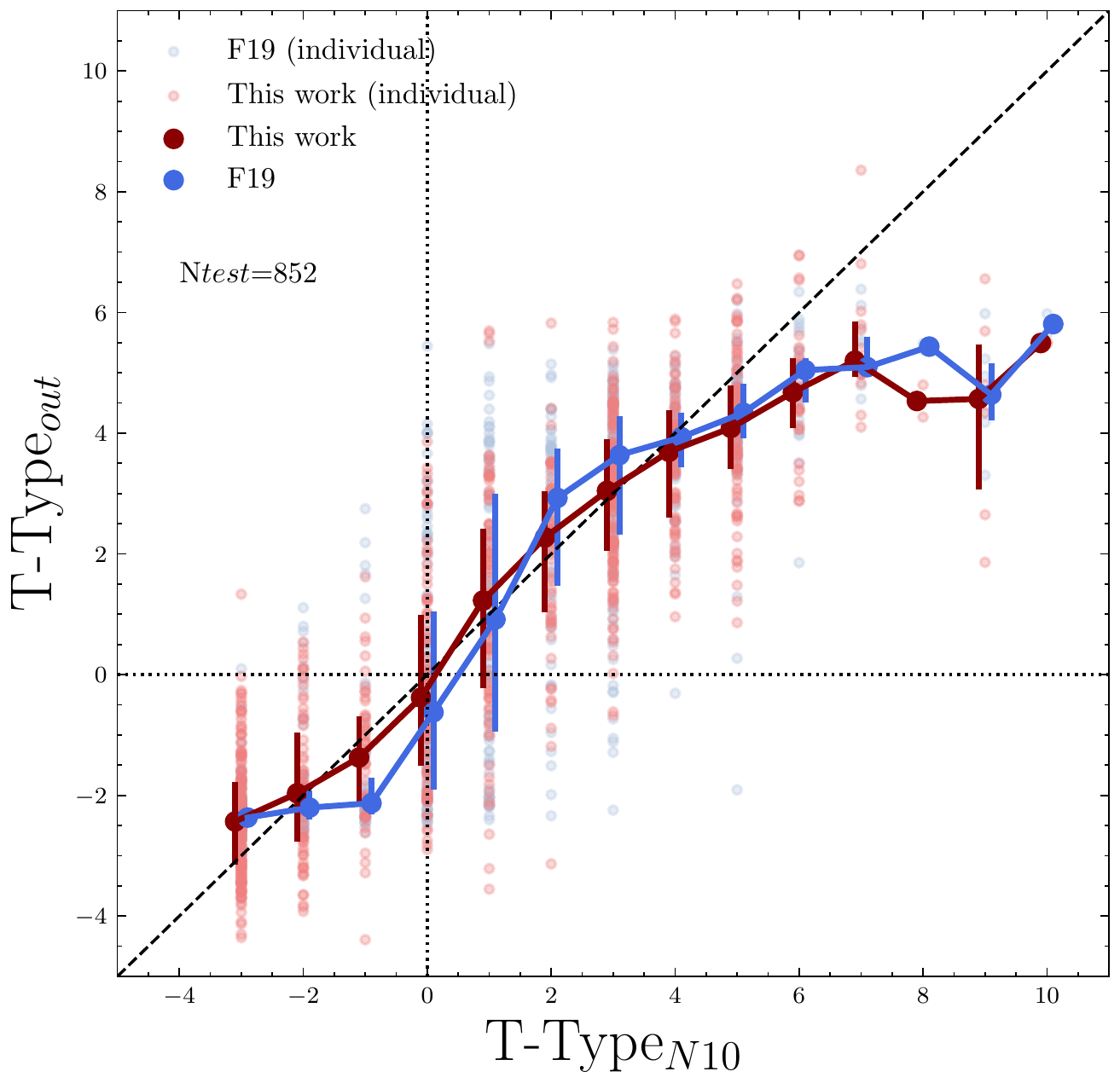}\par 
    \end{multicols}
\caption{Left: Comparison of the T-Type derived from the CNN (the average of 15 models trained with $k$-folding) and the original T-Type (from N10) for the test sample of 1348 galaxies. To better visualise it, we plot average binned values, where the symbol size is proportional to the number of objects in each bin. The red dots (joined by a solid line) show the median value at each T-Type, while the error bars show the inter-quartile ranges (i.e., the difference between 75th and 25th percentiles). The predicted T-Types follow the one to one relation (dashed line) very well up to T-Type $\sim$ 5. Right: Same as the left panel but comparing the new results (red) with the models presented in F19  (blue) for 852 individual galaxies. There is an improvement in the bias, especially at  T-Type~$<$~0 and 1~$<$~T-Type~$<$~4. (The average values and their error bars have been shifted $\pm$ 0.1 T-Types for better visualization.) }  \label{fig:TT}
\end{figure*}

\subsection{Training methodology}
\label{sect:training}

The CNN architecture used for the morphological classifications of the binary models i.e., P$_{\rm LTG}$, P$_{\rm S0}$, P$_{\rm edge-on}$, P$_{\rm bar}$) is identical to that of DS18 (see Figure 1 there for a schematic representation). Namely, the input  are arrays of  dimension (3, 69, 69) and the CNN consists of four convolutional layers with \textit{relu} activation,  filter sizes of 6$\times$6, 5$\times$5, 2$\times$2 and 3$\times$3; 32, 64, 128 and 128 channels; and dropouts of 0.5, 0.25, 0.25 and 0.25, respectively; followed by a fully connected layer of 64 neurons with 0.5 dropout, \textit{sigmoid} activation and \textit{adam} optimizer. The output of the model is one single value, which can be interpreted as the probability of being a positive example. The total number of trainable parameters is 2602849. The CNN was trained for 50 epochs with binary cross-entropy as the loss function.

Due to the complexity of the T-Type classification, we used a slight variation of the CNN architecture described in DS18 to train the T-Type model: the convolutional layers remain as explained above, but the model includes two fully connected layers of 128 and 64 neurons each, with 0.5 dropout. The total number of trainable parameters increases up to  4978657. The CNN was trained for 100 epochs in regression mode  and mean squared error as the loss function.  

One of the main improvements with respect to the MDLM-VAC-DR15 is that the new catalogue includes model uncertainties  provided by the standard deviation obtained with $k$-folding  (with $k=10$, except for the T-Type model where $k=15$). This methodology is very close to deep ensembles -- which are   formally demonstrated to be a Bayesian uncertainty quantification  (see \citealt{Lakshminarayanan2016}) --  and accounts for variations in the initialization of the CNN weights as well as variations due to the training sample. The value reported in the catalogue is the average of the $k$ models and the uncertainty is their standard deviation. This methodology has been demonstrated to improve the performance with respect to the value  of the individual models (see \citealt{VegaFerrero2021}). We reserved a sample which has never gone through the CNN and used it as test sample from which to measure the performance of the $k$-models (see Section \ref{sect:Models}).

\subsection{Models performance}
\label{sect:Models}

In this section we show the performance of the models, i.e., how the predicted morphological classifications compare to the original ones. For this purpose, we define a 'test sample', which is a set of galaxies which were not used to train in any of the $k$-folds. The results shown in this section are obtained by applying the DL models to these test samples.

\subsubsection{T-Type}\label{sect:T-Type}

The T-Type model is a linear regression and the best way to test its performance is to make a one-to-one comparison with the `true' value. Figure \ref{fig:TT} shows there is excellent agreement between the input and predicted T-Types up to T-Type $\sim$ 5, where the predicted T-Type underestimates the correct value. We attribute this to the small number of such objects in the training and test samples (the symbol sizes are proportional to the number of objects in each T-Type bin). In addition, there is also a slight flattening at the lowest T-Type values.  This was evident, and more pronounced, in the older models presented in F19 (as can be seen in the right panel, where the results of the two models are compared) and is the main reason why we also provide a classification between pure ellipticals and lenticulars (P$_{\rm S0}$).  If we limit the analysis to T-Type $<$ 5 the average bias values (T-Type$_{in}$- T-Type$_{out}$) are $b=-0.1$ and $b=-0.4$ for this work and for the F19 models, respectively. This is smaller than typical differences between the visual classifications of different professional astronomers ($b$ $\sim$ $1$). The scatter is also smaller for the new model at intermediate T-Types  (-1, -3) but larger otherwise. In part because, the flattening of the F19 model, specially around T-Type = -2 and T-Type=3, reduces the F19 scatter. 

To summarize: Our new T-Type model shows a smaller bias compared to the one presented in F19 (especially at  T-Type $<$ 0 and 1 $<$ T-Type $<$ 4) and includes an uncertainty value (determined from the standard deviation of the T-Type predicted by each of the 15 models trained with $k$-folding).

\subsubsection{P$_{\rm LTG}$ and P$_{\rm S0}$ models}
\label{sect:P-LTG}

In addition to the T-Type model, the MDLM-VAC-DR17  provides two binary classifications trained with the N10 catalogue. 

The first one, P$_{\rm LTG}$, separates ETGs from LTGs.

The model does an excellent job at separating the E and S2 populations, recovering 98$\%$ of the ellipticals (defined as true negatives; TN, i.e., P$_{\rm LTG} < $  0.5 and labeled as negative in the training sample) and 97$\%$ of the S2 (defined as true positive, TP; i.e., P$_{\rm LTG} > $  0.5 and labeled as positive in the training sample). The separation for the intermediate populations, S0 and S1, is less clean, as expected. For the S0s, 59$\%$ are classified as ETGs and 41$\%$ as LTGs, while for the S1s the fractions of galaxies classified as ETGs and LTGs are 27$\%$ and 73$\%$, respectively. This classification is very useful for making a broad separation between ETGs and LTGs, especially at intermediate T-Types, where the T-Type scatter is large. 

\begin{figure}
    \includegraphics[width=0.9\linewidth]{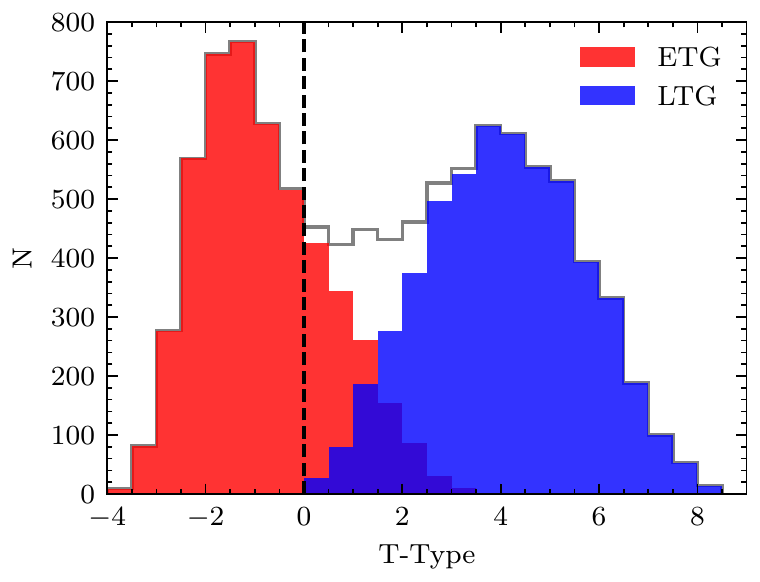}\par 
\caption{Bimodal distribution of the predicted T-Type  is well described by our binary ETG and LTG classifier (recall that LTGs are defined as having P$_{\rm LTG} > 0.5$ while ETGs have P$_{\rm LTG} < 0.5$). The black dashed line at T-Type~=~0 marks the separation between E/S0 and S, based on the T-Type model. Note that there are some galaxies with T-Type~>~0 classified as ETGs, while the opposite is negligible (see discussion in section \ref{sect:P-LTG}).} \label{fig:TT-PLTG}
\end{figure}

Figure \ref{fig:TT-PLTG} compares our T-Type and P$_{\rm LTG}$ predictions for the full MaNGA DR17. The bimodality distribution of T-Types is very well traced by the ETG and LTG populations. Only 2 galaxies with T-Type $<$ 0 are classified as LTG (P$_{\rm LTG}$ > 0.5). On the other hand, 1315 galaxies with T-Type $>$ 0 and are classified as ETGs (P$_{\rm LTG}$ < 0.5).  

\begin{figure}
    \includegraphics[width=0.9\linewidth]{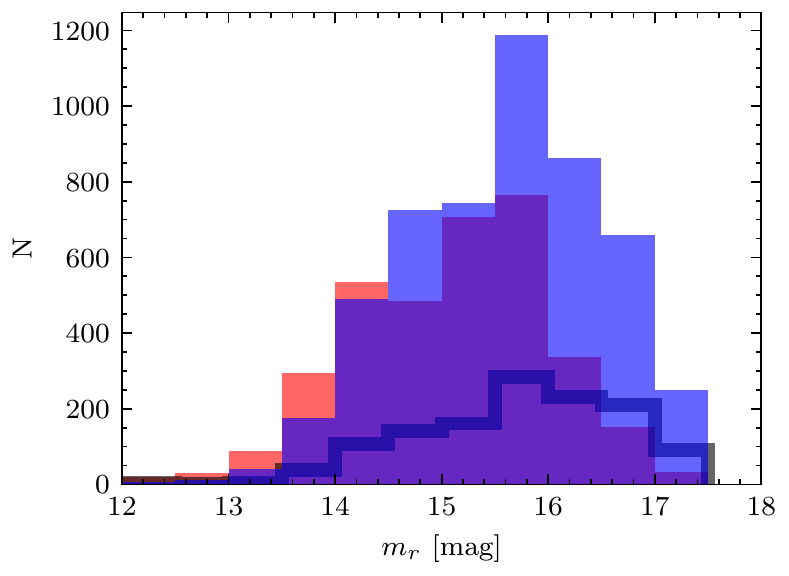}\par 
    \includegraphics[width=0.9\linewidth]{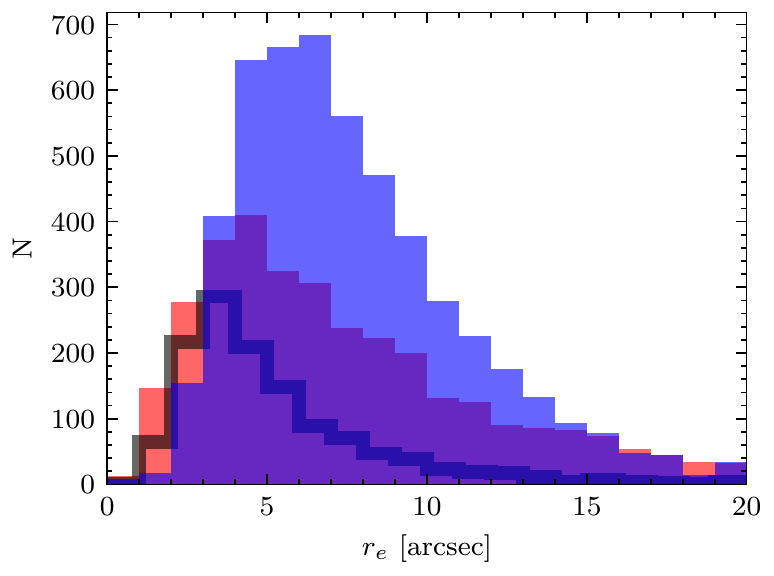}\par 
    \includegraphics[width=0.9\linewidth]{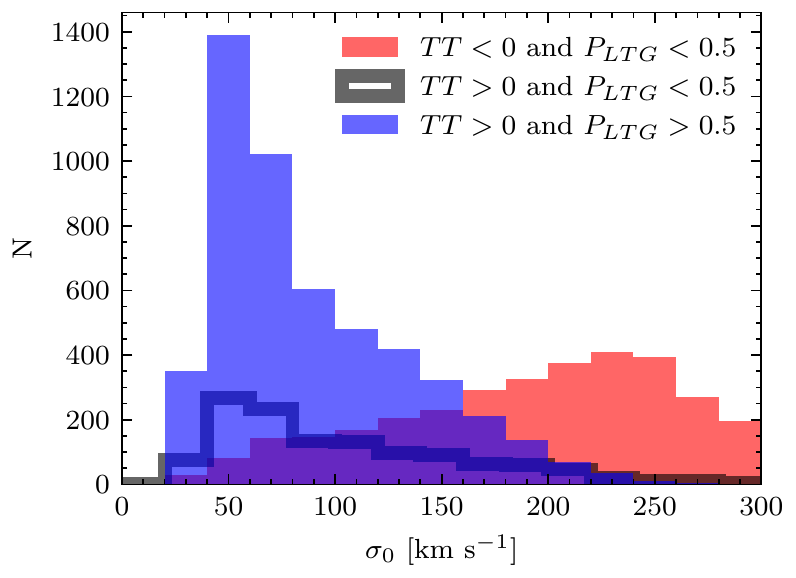}\par 
\caption{Galaxies having inconsistent T-Type and P$_{\rm LTG}$ classifications (empty histograms) tend to be faint (top), have small angular sizes (middle) and small central velocity dispersions (bottom).  } \label{fig:TT-PLTG_bad}
\end{figure}

Figure \ref{fig:TT-PLTG_bad} shows the distribution in apparent magnitude, angular size and central velocity dispersion\footnote{Defined as the velocity dispersion at 0.25 arcsec  derived from the MaNGA data-analysis pipeline (following the methodology described in \citealt{DS2019}).}, of the galaxies having inconsistent T-Type and P$_{\rm LTG}$ classifications. These galaxies (empty histograms) occupy the faint end of the magnitude distribution (top), have small angular sizes (middle) and low central velocity dispersions (bottom) (similar to the LTGs). I.e., they are probably too faint or small to clearly show spiral structure. We conclude that these galaxies are the most difficult to classify: the T-Type classification might be correct, while the P$_{\rm LTG}$ model is actually separating galaxies with evident spiral features from galaxies which look smoother. An example of that kind of galaxies can be seen in the top cutout of figure \ref{fig:TT1}.

\begin{figure}
    \includegraphics[width=0.9\linewidth]{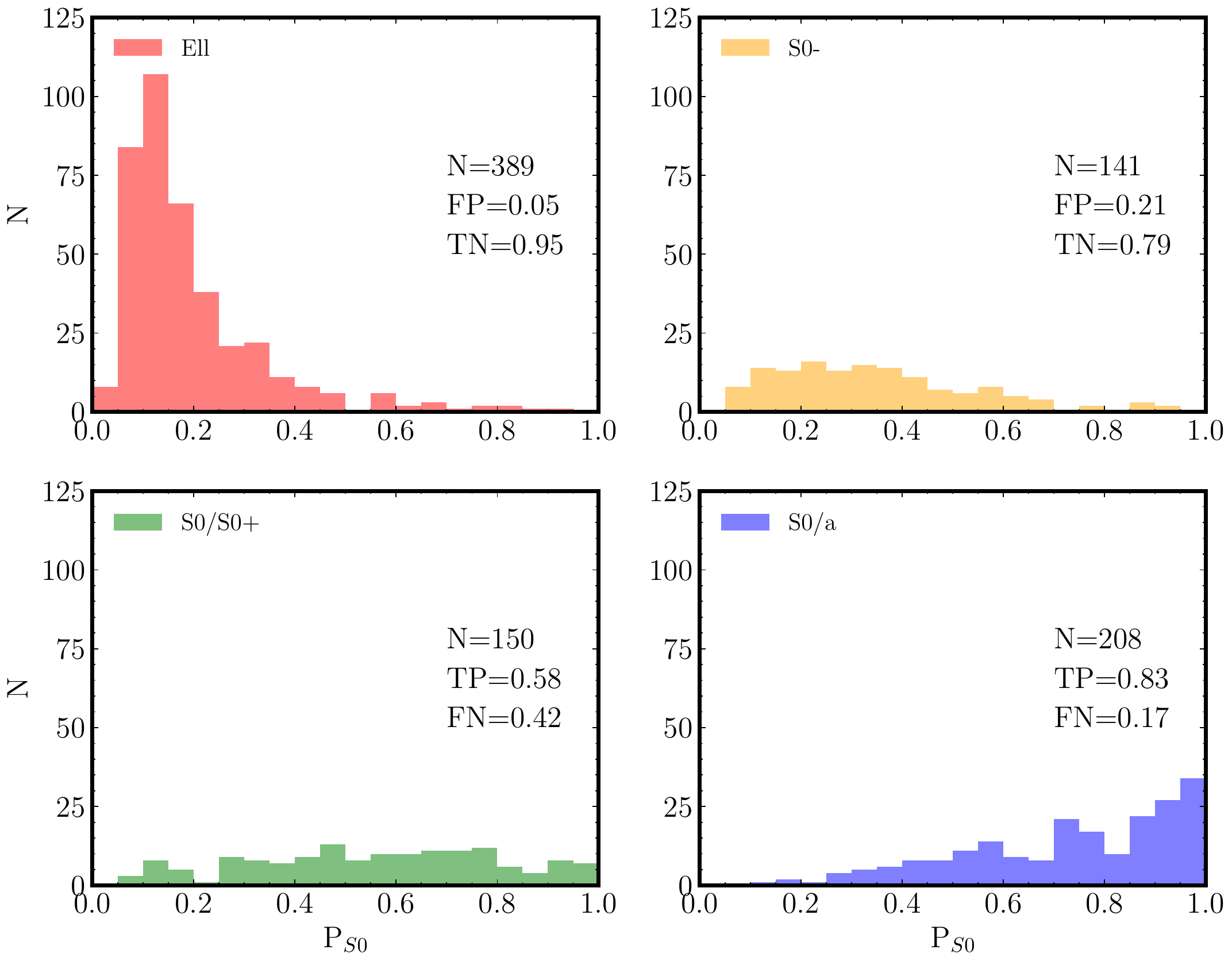}\par 
\caption{Distribution of P$_{\rm S0}$ separating Es from S0s for different classes from N10.  Each panel also provides the total number of galaxies and the false positive and true negative rates (top) or the true positive and false negative rates (bottom).} \label{fig:PS0}
\end{figure}

The second binary classification trained with the N10 catalogue separates S0s from pure ellipticals (E). This model, P$_{\rm S0}$, is trained with galaxies having T-Type $<$ 0 (from N10) and therefore, is only meaningful for galaxies with negative values of the predicted T-Type.  The reason for constructing this model is, again, the large scatter around intermediate T-Types, where the transition between Es and S0s occurs. Figure \ref{fig:PS0} shows that, in the test sample, the model classifies as ellipticals 95$\%$ of the  galaxies from N10 with T-Type=-5 and as S0 83$\%$ of the galaxies classified as S0/a from N10 (with a T-Type=0).  The predicted P$_{S0}$ for the galaxies with T-Types in between is distributed around intermediate values, as expected. The performance of this P$_{\rm S0}$ is not as good as P$_{\rm LTG}$, which is reasonable given that what separates Es from S0s is rather subtle compared to the differences between ETGs and LTGs.  

Hereafter -- as done in F19 --, we classify the galaxies into three broad categories (E, S0 and S) by combining the T-Type and P$_{\rm S0}$ as follows:

\begin{itemize}
    \item E: T-Type $<$ 0 and P$_{\rm S0}$ $<$ 0.5 
    \item S0: T-Type $<$ 0 and P$_{\rm S0}$ $>$ 0.5
    \item S: T-Type $>$ 0  
\end{itemize}

In some sections we further subdivided the S galaxies in two sub-samples:
\begin{itemize}
    \item S1: 0 $<$ T-Type $<$ 3  
    \item S2: T-Type $>$ 3  
\end{itemize}

\subsubsection{Visual classification}
To have a more robust classification, and since the sample size allows it, we have also carried out a visual inspection of all the galaxies.  The catalogue includes two columns reporting the results:

 (i) Visual Class, which corresponds to the visual classification assigned to each galaxy (VC=1 for elliptical, VC=2 for S0, VC=3 for S/Irr and VC=0 for unclassifiable);

\begin{figure}
    \includegraphics[width=0.9\linewidth]{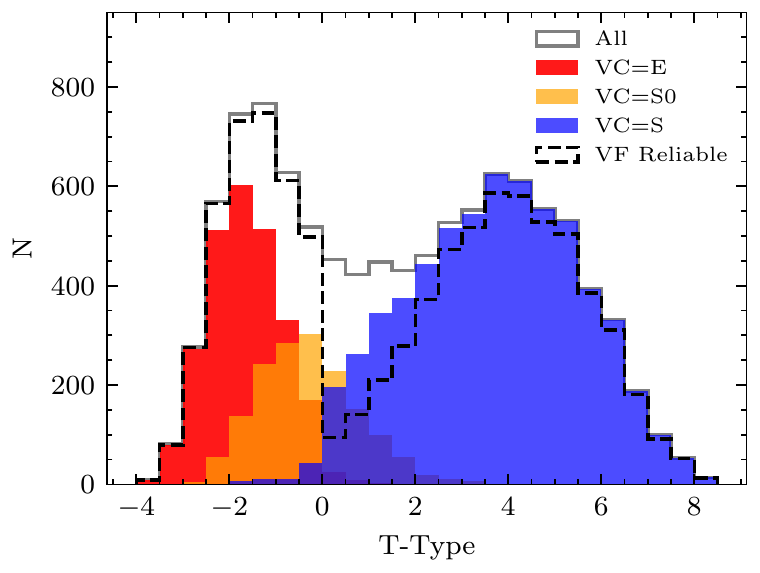}\par 
\caption{Predicted T-Types for galaxies according to their visual classification. The black dashed line shows  galaxies with certain visual classifications (VF=0). There is an evident drop of certain visual classifications for the intermediate T-Types ($\sim$ 0), where the distinction between E/S0 and S is very subtle.} \label{fig:VC}
\end{figure}

 (ii) Visual Flag, which reports the level of confidence in our visual class (VF=0  for reliable and VF=1 for uncertain classifications).\\
The visual classifications were based on the models (i.e., they were not blind) to spot evident miss-classifications.
Figure~\ref{fig:VC} shows that the visual classifications correlate very well with the predicted T-Types. Galaxies with VC=1 (E) peak around T-Type $\sim$ 2 and barely extend beyond T-Type $>$ 0, galaxies with VC=3 (S) peak around T-Type $\sim$ 4 and barely extend below T-Type $<$ 0, while galaxies with VC=2 (S0) tend to have intermediate T-Types with a tail which extends to T-Type $>$ 0. Thus, the reader should be aware that selecting a sample of S0 galaxies  based on T-Type (i.e. T-Type < 0 and P$_{ \rm S0}$ > 0.5) produces a pure but not complete sample.  It also shows that galaxies with low confidence visual classifications  are mostly those with intermediate T-Types.

To quantify the comparison, Table \ref{tab:VC} shows the number of galaxies classified as E, S0 or S according to the combination of the T-Type and P$_{\rm S0}$ values (as defined in section \ref{sect:P-LTG}) and the visual classification. Only 5$\%$ of galaxies with elliptical morphologies (according to the models) are visually classified as S0 (4$\%$) or S (1$\%$). Similarly, for the galaxies classified as S0, there is 96$\%$ agreement, with most of the discrepancies coming from galaxies visually classified as S. The more important miss-match is for the S sample, where 9$\%$ of galaxies classified as S by the models are assigned type S0 after visual inspection. Note that, since the P$_{\rm S0}$ classification is only meaningful for galaxies with T-Type $<$ 0, there is no `model' to distinguish between S0 and S for galaxies with T-Type $>$ 0.

We also note that 1/3 of the galaxies  with T-Type > 0 and VC=2 (S0) have a large probability of being edge-on (P$_{\rm edge-on} > 0.5$, see section \ref{sect:edge-bar}), which explains the discrepancies (it is almost impossible to distinguish an S from a S0 when seen edge-on). In fact, if we focus on galaxies with reliable visual classifications (VF=0), the agreement is significantly improved up to 99$\%$ (see upper panel with VF=0 in Table \ref{tab:VC}).

\begin{table}
\caption{Comparison of automated (by combining the T-Type and P$_{\rm S0}$ models or according to P$_{\rm LTG}$) and visual  classifications (VC=1 for elliptical, VC=2 for S0, VC=3 for S/Irr and VC=0 for unclassifiable). In the left most column we report the number of galaxies of each type, while the other columns show the percentage of each type with the corresponding visual classification for all galaxies (top) and for galaxies with reliable visual classifications (VF=0, bottom).}
\label{tab:VC}
\begin{tabular}{lrcccc}
\hline
    TT+P$_{\rm S0}$& & VC$\_$E  & VC$\_$S0 & VC$\_$S & VC$\_$unc\\
   & &  (VC=1)  & (VC=2) & (VC=3) & (VC=0)\\  
   \hline  
   & All &   &   &  & \\    
\hline
 E &  2632  & 95  & 4  & 1 & $<$1 \\
 S0 & 963   & $<$ 1  & 96  &4  &  $<$1  \\
 S & 6698  & $<$ 1  &  9 & 90  &  $<$ 1 \\
\hline
   & VF=0 &   &   &  & \\    
\hline
 E &  2598 & 95  & 4  & 1 & $<$1 \\
 S0 & 922 & $<$ 1  & 97 & 3 &  $<$1  \\
 S & 5320 & $<$ 1  &  1 & 99 &  $<$ 1 \\
\hline
\hline

  P$_{\rm LTG}$& &  VC$\_$E  & VC$\_$S0 & VC$\_$S & VC$\_$unc\\
    & &  (VC=1)  & (VC=2) & (VC=3) & (VC=0)\\ 
 \hline
   &  All & &  & & \\
\hline
 ETG &  4908  & 52 & 32 & 16 & $<$ 1\\
 LTG & 5385  &  0  & $<$ 1 & 99 & $<$ 1\\
\hline
 &  VF=0 &   &     &  & \\  
\hline
 ETG &  3700 & 67& 27 & 5 & $<$ 1\\
 LTG & 5140 &  0  & $<$ 1 & 99 & $<$ 1\\
\end{tabular}
\end{table}

The bottom part of Table \ref{tab:VC} compares the visual classifications with the separation between ETG and LTG according to the P$_{\rm LTG}$ model. The LTG sample is very pure:  99$\%$ of the galaxies with P$_{\rm LTG} > 0.5$ were visually classified as S. On the other hand, the ETG population (P$_{\rm LTG} <= 0.5$) is composed of 52$\%$ Es, 32$\%$ S0s and 16$\%$ Ss, according to the visual classification.  Considering only galaxies with reliable visual classifications, the fractions of E increases up to 67$\%$, while the fraction of S0 and S become 27 and 5$\%$ respectively. 

Interestingly, the percentage of galaxies classified as ETG with reliable visual classifications is only 75$\%$ (3700/4908), comparable to the same fraction for galaxies classified as S (5320/6698=79$\%$). While this may seem contradictory, it is because most of the galaxies with VF=1 are those with T-Type > 0 (classified as S by the T-Type models) and P$_{\rm LTG}$ < 0.5 (classified as ETGs by the P$_{\rm LTG}$ model), i.e., they are the faint and small galaxies difficult to classify and for which the T-Type and the P$_{\rm LTG}$ classifications disagree: 86$\%$ of the galaxies with T-Type $> 0$ and  P$_{\rm LTG}$ $< 0.5$ have VF=1 (1134/1315) and 78$\%$ of galaxies with VF=1 have T-Type $> 0$ and  P$_{\rm LTG}$ $< 0.5$ (1134/1453).

\begin{figure}
    \includegraphics[width=0.9\linewidth]{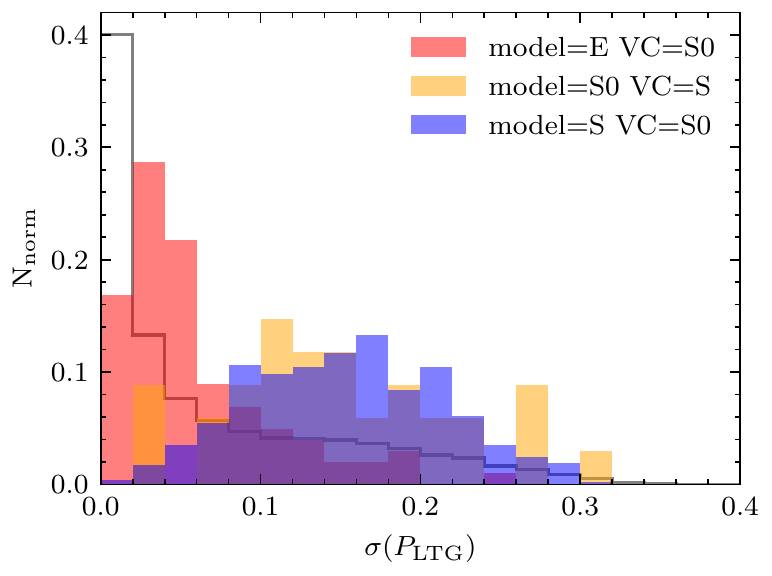}
    \caption{P$_{\rm LTG}$ uncertainty for the full sample (empty grey histogram) and for galaxies whose visual classification is in disagreement with the classification obtained by combining the T-Type and P$_{\rm S0}$ models (red for Es visually classified as S0, orange for S0s visually classified as S and blue for Ss visually classified as S0 - the other combinations are not shown due to their small number, as detailed in Table \ref{tab:VC}.) The uncertainties for the `miss-classified' galaxies are significantly larger.} \label{fig:Sigma-P_LTG}
\end{figure}

To test weather the model uncertainties correlate with the `miss-classifications',  figure \ref{fig:Sigma-P_LTG} shows the  standard deviation of the value returned by $k= 10$  models separating ETGs from LTGs for galaxies whose visual classification is different than the classification obtained by combining T-Type and P$_{\rm S0}$ models. The uncertainties are significantly larger than for the overall population. The same is true for the  P$_{\rm S0}$ uncertainties for Es visually classified as S0s (not shown here).

On the contrary, the `miss-classified' galaxies do not show larger T-Type uncertainties than the full sample. Although that might seem unexpected, we must take into account that the T-Type model is a linear regression and is not aware of our `artificial' separation between E/S0s and S at T-Type = 0 defined in section \ref{sect:P-LTG}. What happens for these galaxies is that they are close to that limit, with average  values of T-Type=-0.3 and T-Type=0.67 for the `miss-classified' S0 and S, respectively. To quantify this uncertainty, we have generated 100 classifications based on the average T-Type by bootstrapping one of the $k=15$ models in each realization. The percentage of galaxies which changes class  (i.e., has median T-Type $>$ or $<$ 0 in a different realization) more than 10 times is $\sim$ 5\% for the overall population, while this happens for 40$\%$ of the `miss-classified' galaxies, demonstrating that the T-Type scatter is consistent with these sub-samples being more difficult to separate into the broad E/S0 and S classes.

\begin{figure}
    \includegraphics[width=0.9\linewidth]{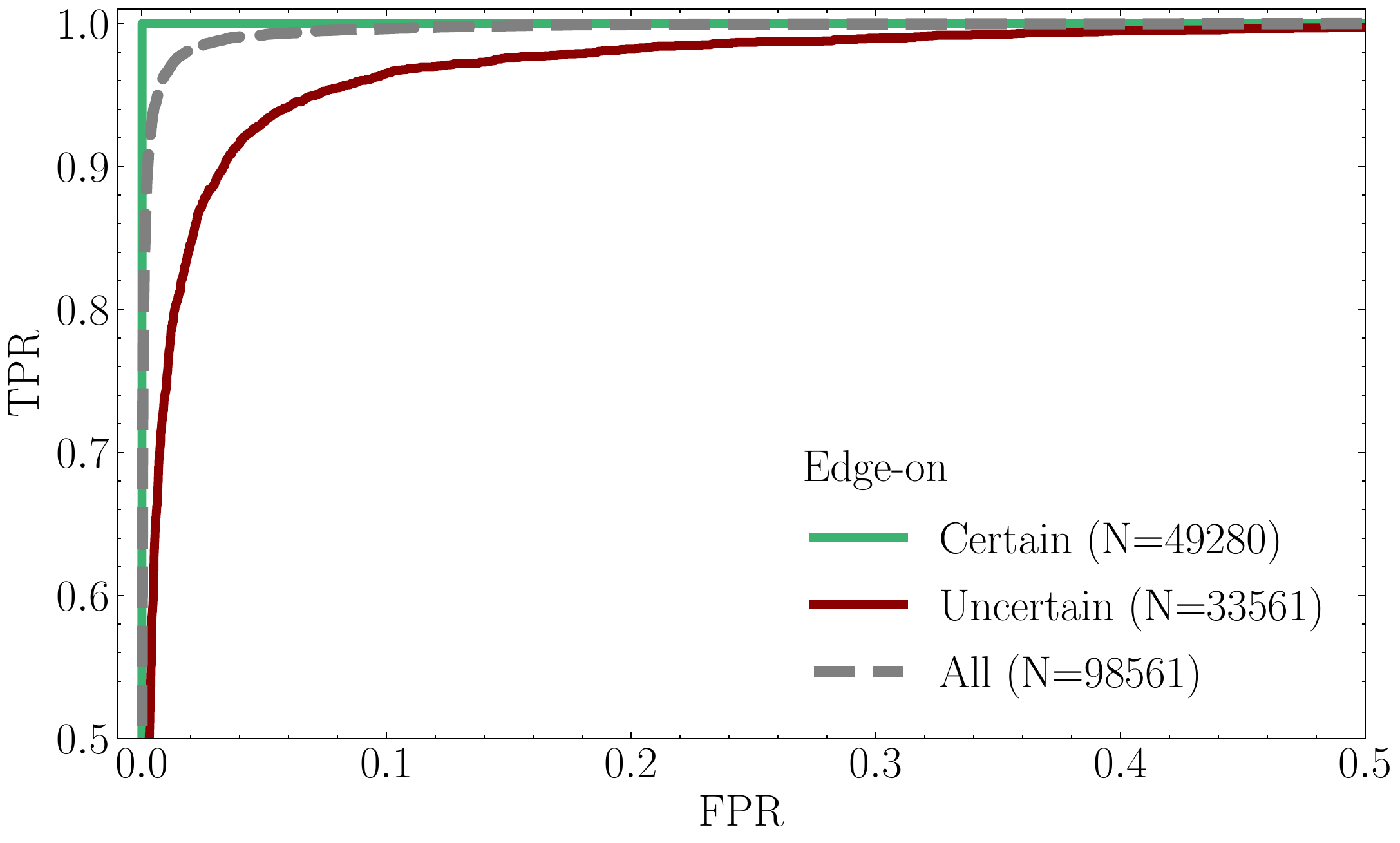}
    \includegraphics[width=0.9\linewidth]{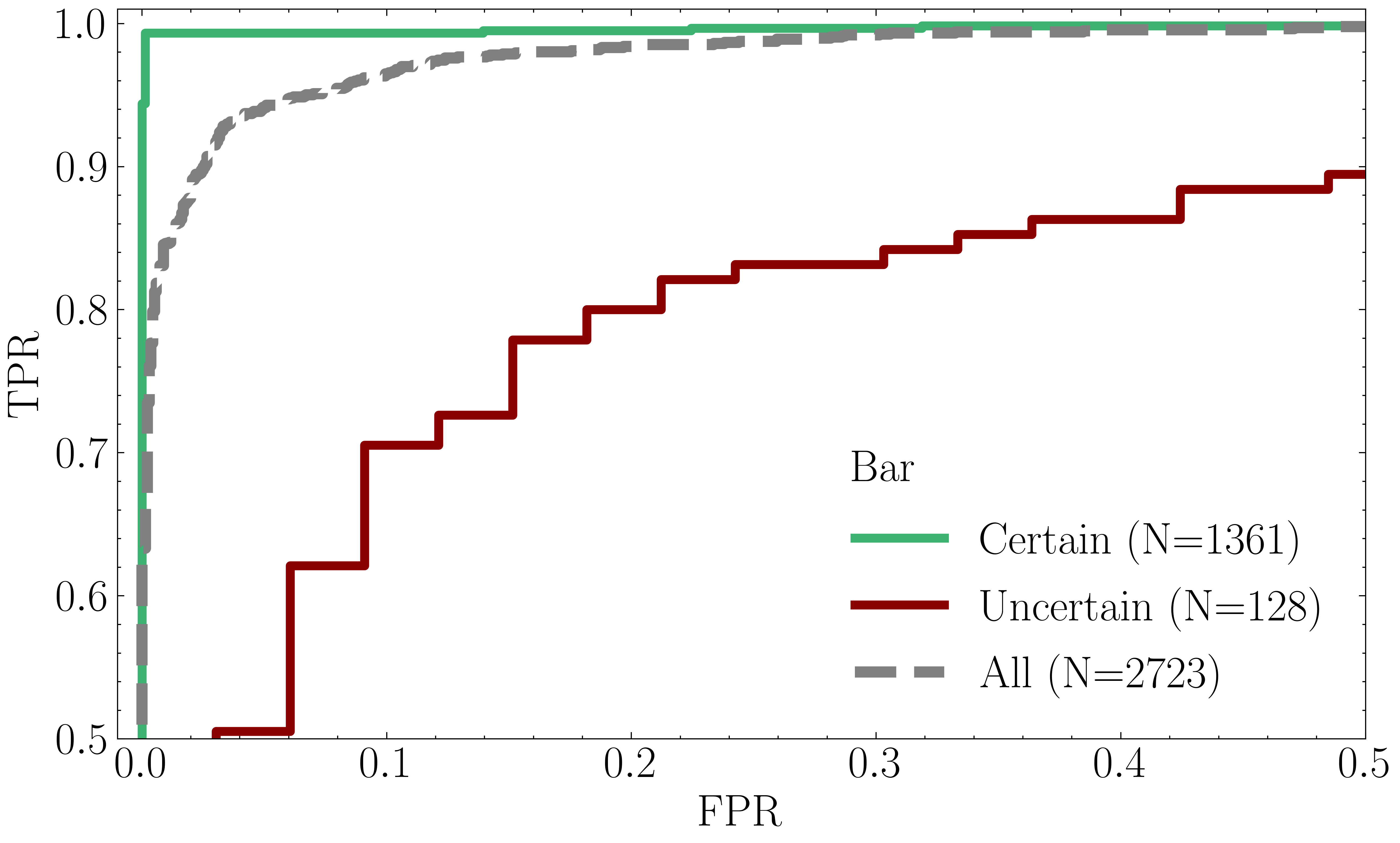}
    \caption{ROC curve -- True Positive Rate (TPR) versus False Positive Rate (FPR) -- for the edge-on (top) and bar (bottom) classifications.  Gray, green and red curves show the full test sample, the subset for which the model uncertainties are below the average, and the subset for which the uncertainties are larger than 3$\sigma$, with $N$ representing the size of each sub-sample. Galaxies with more certain classifications (green) show better performance. } \label{fig:ROC-edge}
\end{figure}

\subsubsection{Edge-on and bar classifications}
\label{sect:edge-bar}

\begin{table*}
\caption{Accuracy, precision, recall and F1 score for P$_{\rm edge-on}$ and P$_{\rm bar}$, as well as the number of galaxies in the test sample and the fraction of those labeled as positive (according to GZ2). }
\label{tab:Edge-bar}
\begin{tabular}{lcccccc}
\hline
 Model & N$_{test}$ & $\%$ Positives & Accuracy & Precision & Recall & F1 \\
\hline
 P$_{\rm edge-on}$ &  98561 & 14 & 0.98   & 0.87  & 0.98  & 0.93  \\
 P$_{\rm bar}$     &  2723 & 50 & 0.93 & 0.92 & 0.90 &  0.93 \\
\hline
\end{tabular}
\end{table*}

\begin{table*}
\centering
MDLM-VAC: The MaNGA Deep Learning Morphological VAC
\begin{tabular}{ |p{1.7cm}|p{1cm}|p{12.5cm}| }
\hline 
Column Name & Data Type & Description  \\ 
\hline 
INTID &	int &	Internal identification number \\
MANGA-ID &	string	& MaNGA identification  \\
PLATEIFU &	string &	MaNGA PLATE-IFU \\
OBJID	 & long64 & 	SDSS-DR14 photometric identification number \\
RA	 & double & 	Object right ascention (degrees) \\
DEC& 	double & 	Object declination (degrees) \\
Z	& double & 	NSA redshift (from SDSS when not available) \\
DUPL$\_$GR&  	int & 	Group identification number for a galaxy with multiple MaNGA spectroscopic observations \\
DUPL$\_$N & 	int	&  Number of multiple MaNGA spectroscopic observations associated with DUPL$\_$GR \\
DUPL$\_$ID & 	int & 	Identification number of the galaxy in the group DUPL$\_$GR \\
TType & 	double & 	T-Type value trained with the N10 catalogue. The value in the catalogue is the average of 15 $k$-fold models. TType < 0 for ETGs. TType > 0 for LTGs. \\
TT$\_$std & 	double & 	Standard deviation of the value returned by the $k$=15 T-Type models. Can be used as a proxy of the T-Type uncertainty. \\
P$\_$LTG & 	double	&  Probability of being LTG rather than ETG. Trained with the N10 catalogue. \\
P$\_$LTG$\_$std	 &  double & 	Standard deviation of the value returned by the $k$=10  P$_{\rm LTG}$ models. Can be used as a proxy of the  P$_{\rm LTG}$  uncertainty. \\
P$\_$S0 & 	double	&  Probability of being S0 rather than pure elliptical, trained with the N10 catalogue. Only meaningful for galaxies with T-Type <=0 and not seen edge-on. \\
P$\_$S0$\_$std & 	double & 	Standard deviation of the value returned by the $k$=10  P$_{\rm S0}$ models. Can be used as a proxy of the  P$_{\rm S0}$ uncertainty. \\
P$\_$edge-on	&  double & 	Probability of being edge-on, trained with the GZ2 catalogue. \\
P$\_$edge-on$\_$std & 	double & 	Standard deviation of the value returned by the $k$=10  P$_{\rm edge-on}$ models. Can be used as a proxy of the  P$_{\rm edge-on}$ uncertainty. \\
P$\_$bar	& double & 	Probability of having a bar signature, trained with GZ2 catalogue. Edge-on galaxies should be removed to avoid contamination. \\
P$\_$bar$\_$std	& double & 	Standard deviation of the value returned by the $k$=10 P$_{\rm bar}$ models. Can be used as a proxy of the  P$_{\rm bar}$ uncertainty. \\
Visual$\_$Class & 	int & 	Visual classification: VC=1 for ellipticals, VC=2 for S0, VC=3 for S (including Irregulars), VC=0 for unclassifiable. \\
Visual$\_$Flag & 	int & 	Visual classification flag: VC=0 certain visual classification, VC=1 uncertain visual classification. \\

\hline 
\hline 
\end{tabular}
\caption{Content of the Deep Learning Morphological catalog for the DR17 MaNGA sample. This catalog  is available online$^7$. } 
\label{tab:morph}
\end{table*}

\begin{table*}
\centering
FRACTION OF GALAXIES\\
\begin{tabular}{lccccc}
  \hline
   & Good fits (Ser and/or SerExp) &  Good fits  &  Good Ser and SerExp fits   & Good Ser fits  & Good SerExp fits  \\
    & &  $\&$ Reliable Visual Class &   $\&$ Reliable Visual Class  & $\&$ Reliable Visual Class &  $\&$ Reliable Visual Class \\  
 
   & (FLAG$\_$FIT $\ne 3$) &   (FLAG$\_$FIT  $\ne 3$ + VF=0) &  (FLAG$\_$FIT = 0  + VF=0) & (FLAG$\_$FIT = 1  + VF=0) & (FLAG$\_$FIT = 2  + VF=0) \\

  \hline
  Class &  & & Based on T-Type + P$_{\rm S0}$  & &  \\
     \hline
  E & 0.968 & 0.956 &  0.182&  0.618 &  0.200 \\
  S0 & 0.964 & 0.930 &  0.123 & 0.450 & 0.427 \\
  S1 & 0.966 & 0.550 & 0.077 & 0.390 & 0.532 \\
  S2 & 0.966 & 0.920 & 0.048 & 0.640 & 0.312 \\
    \hline
&   &  & Based on Visual Classification  & &  \\
     \hline
  E & 0.966 & 0.946 & 0.184 &  0.630 & 0.186 \\
  S0 & 0.973 & 0.626 & 0.125 &  0.438 & 0.437 \\
  S1 & 0.966 & 0.692 & 0.078 & 0.392 & 0.529 \\
  S2 & 0.966 & 0.921 & 0.048 & 0.640 &  0.312 \\
  \hline
    &   & & Based on P$_{\rm LTG}$  & &    \\
     \hline
  ETG & 0.967 & 0.729 & 0.161 & 0.561  & 0.277 \\
  LTG & 0.966 & 0.925 & 0.056 & 0.575 & 0.369 \\

  \hline
  \hline
\end{tabular}
\caption{Left most columns: Fraction of galaxies of a given morphological type which have PyMorph parameters (from S{\'e}rsic and/or SerExp, i.e. FLAG$\_$FIT $\ne$ 3) and reliable visual classification (VF=0). Right most columns: fraction of galaxies with FLAG$\_$FIT  $\ne 3$, VF=0 and having 2 components (FLAG$\_$FIT = 2), 1 component (FLAG$\_$FIT = 1), or for which both descriptions are equally acceptable (FLAG$\_$FIT = 0). }
\label{tab:fit}
\end{table*}

The catalogue includes two binary classifications based on the GZ2 catalogue \citep{Willett2013}: identification of edge-on galaxies (P$_{\rm edge-on}$) and identification of galaxies with bar signatures (P$_{\rm bar}$). The Receiver Operating Characteristic curve (ROC) --  the True Positive Rate (TPR) versus False Positive Rate (FPR) -- is commonly used to assess the performance of binary classifications. Figure~\ref{fig:ROC-edge} shows the ROCs for P$_{\rm edge-on}$ and P$_{\rm bar}$. The models perform well, with accuracy of 98 and 93$\%$, respectively. The precision, recall and F1 scores,  defined as \\

\indent Precision = TP/(TP+FP) \\
\indent Recall = TP/(TP+FN ) \\
\indent F1 Score = 2$\times$(Recall $\times$ Precision) / (Recall + Precision)\\

\noindent are given in Table~\ref{tab:Edge-bar}. These values reach 100$\%$ in all cases when computed for the sub-sample of galaxies with certain classifications, defined as those with standard deviations of the predicted models below the average (see also green lines in Figures \ref{fig:ROC-edge}). This is reassuring, not just regarding the quality of our classifications, but also on the meaning of the reported model uncertainties.

Combining these classifications with the previous  ones, only 9  galaxies classified as E (according to T-Type and P$_{\rm S0}$)  have P$_{\rm edge-on}> $  0.8 (compared to  1551 for the full catalogue), while 53  have P$_{\rm bar} >$  0.8 (compared to 1300 for the full catalogue). The number of ETGs ( i.e., P$_{\rm LTG} <$  0.5) with P$_{\rm edge-on}$ or P$_{\rm bar} > $  0.8 is 224 and 246, respectively (out of 4908 ETGs).  These small numbers are expected, because E galaxies should not have bar features or disk shapes. We want to highlight that it is very difficult to distinguish between S0 and S galaxies when seen edge-on, and therefore, the separation between these two families for galaxies with large  P$_{\rm edge-on}$ is not accurate. In fact, one third of the galaxies with T-Type $>$ 0 and VC=S0 have  P$_{\rm edge-on} > 0.5$ (see discussion related to Table \ref{tab:VC}).

\subsection{The MDLM-VAC-DR17 catalog}
Table~ \ref{tab:morph} shows the format of the MDLM-VAC-DR17 catalog. The catalogue is released with the SDSS DR17 and is available online \footnote{www.sdss.org/dr17/data\textunderscore access/value-added-catalogs/?vac\textunderscore id=manga-morphology-deep-learning-dr17-catalog}. It includes the classifications discussed in the previous sections plus additional information for the galaxies, such as their coordinates, redshift or duplicates.

In contrast to the DR15 version, MDLM-VAC-DR17 does not include P$_{\rm disk}$ or P$_{\rm bulge}$ values since the B/T and $b/a$ values from the MPP-VAC are sufficient for providing such estimates. In addition, MDLM-VAC-DR17 no longer reports P$_{\rm merger}$ because  it does not properly identify true (3D) mergers but rather projected neighbours, (also see discussion in DS18).

\subsubsection{A note on the selection of S0s}

There are several ways to combine the morphological classifications provided in the MDLM-VAC-DR17 to construct samples of E, S0 and S galaxies. Depending on the scientific purpose, users can be more (or less) restrictive in order to obtain more pure (or complete)  samples. The more restrictive selection would be to combine all the information included in the catalogue as follows:

\begin{itemize}
    \item E: (P$_{\rm LTG}$ < 0.5) and (T-Type < 0) and (P$_{\rm S0}$ < 0.5) and (VC=1) and (VF=0)
    \item S0: (P$_{\rm LTG}$ < 0.5) and (T-Type < 0) and (P$_{\rm S0}$ > 0.5) and (VC=2) and (VF=0)
    \item S: (P$_{\rm LTG}$ > 0.5) and (T-Type > 0) and (VC=3) and (VF=0)    
\end{itemize}

This selection will return 2467, 891 and  5125 galaxies classified as E, S0 and S, respectively. However, there would be 1810 galaxies ($\sim$~18$\%$ of the sample) which do not belong to any of the classes.

If the selection is based on the reliable visual classifications (i.e., VF=0), there will be 2474 Es, 1031 S0s and 5325 Ss. But again, there is a large fraction of galaxies ($\sim$~14$\%$) without a class.

Alternatively, the classification could be based on the combination of P$_{\rm LTG}$ and P$_{\rm S0}$ (which selects 2774 Es, 2134 S0 and 5385 Ss) or on the combination of T-Type and P$_{\rm S0}$ (which selects 2632 Es, 963 S0 and 6698 Ss). It is evident that the most affected sample by the classification criteria are the S0 galaxies.  Out of the 1315 galaxies with T-Type > 0 and P$_{\rm LTG}$ < 0.5,  541 (44$\%$) are visually classified as S0s but only 33 of these have reliable visual classifications (VF=0). As already noted throughout the text (see section \ref{sect:P-LTG}), S0 galaxies are sometimes very difficult to distinguish from Sa and it is practically impossible to identify them when seen edge-on. Therefore, we strongly recommend that catalogue users test the effects different selection criteria may have on their scientific conclusions, especially when dealing with S0s.

\begin{figure*}
    \includegraphics[width=0.3\linewidth]{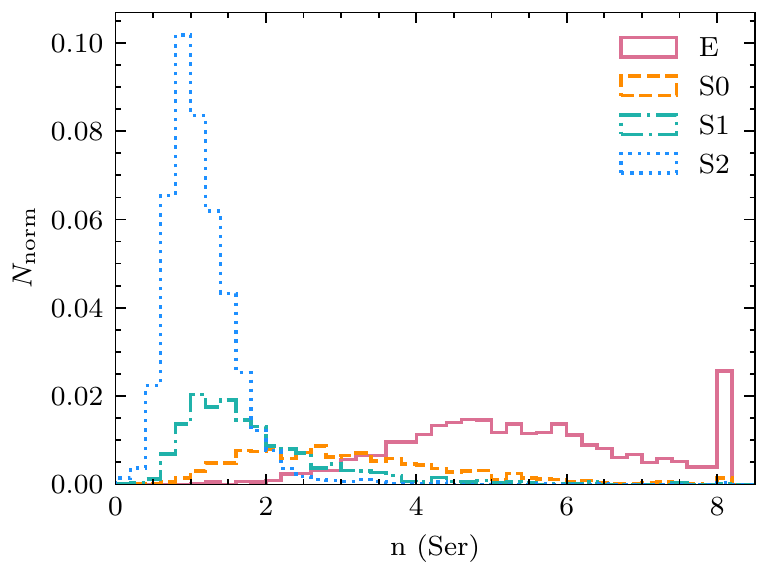}
    \includegraphics[width=0.3\linewidth]{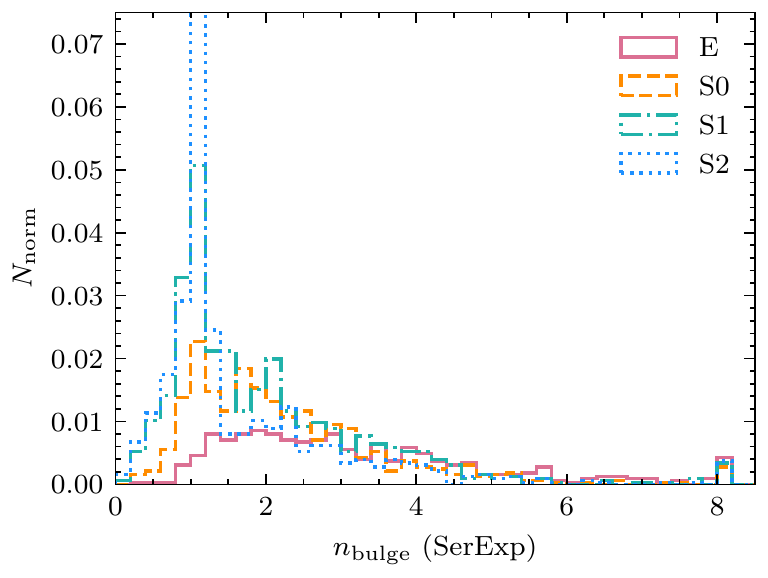}   
    \includegraphics[width=0.3\linewidth]{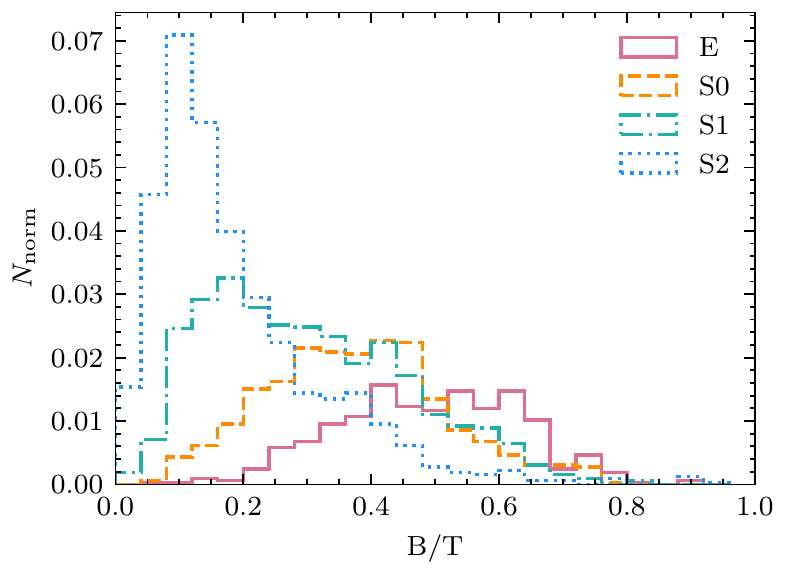} 
    \caption{Distribution of $r$-band $n$ and $n_{\rm bulge}$ for galaxies better described by a 1-component fit (FLAG$\_$FIT=1, left), and 2-components (FLAG$\_$FIT=2, middle). For galaxies with FLAG$\_$FIT=2, also shown (right) the distribution of B/T. Galaxies are color coded by morphology based on the VC values, with an additional separation between S1 and S2 at  T-Type=3. There is a strong correlation between morphological classes and photometric parameters. (The peak height of the S2 histogram is 0.153.) }
    \label{fig:nBT_flag_fit}
\end{figure*}

\begin{figure}
    \includegraphics[width=0.9\linewidth]{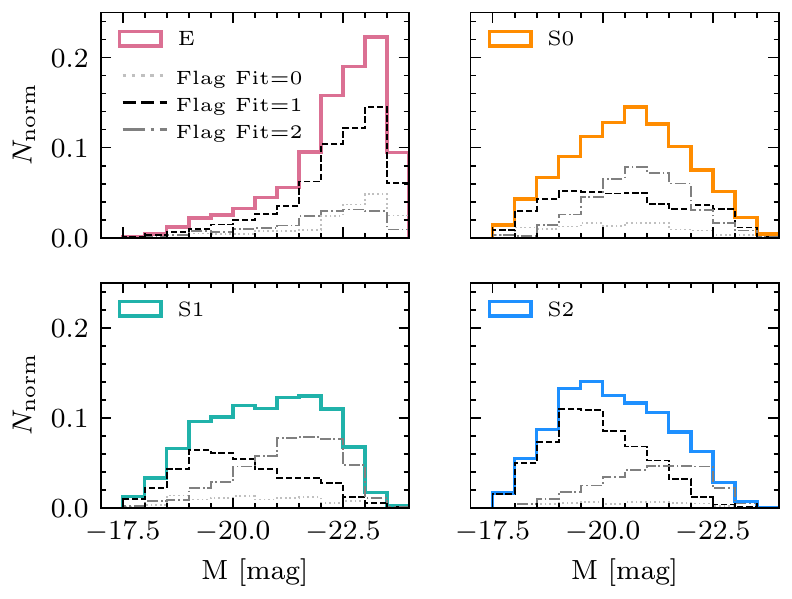}\par 
\caption{Distribution of $r$-band absolute magnitude for galaxies selected on the basis of their morphology and FLAG$\_$FIT.  Es are the brightest, while S peak at fainter absolute magnitudes, especially those with preferred 1-component fit (FLAG$\_$FIT=1 -- i.e., without a bulge component). 
} \label{fig:Mr_flag_fit}
\end{figure}

\begin{figure}
    \includegraphics[width=0.9\linewidth]{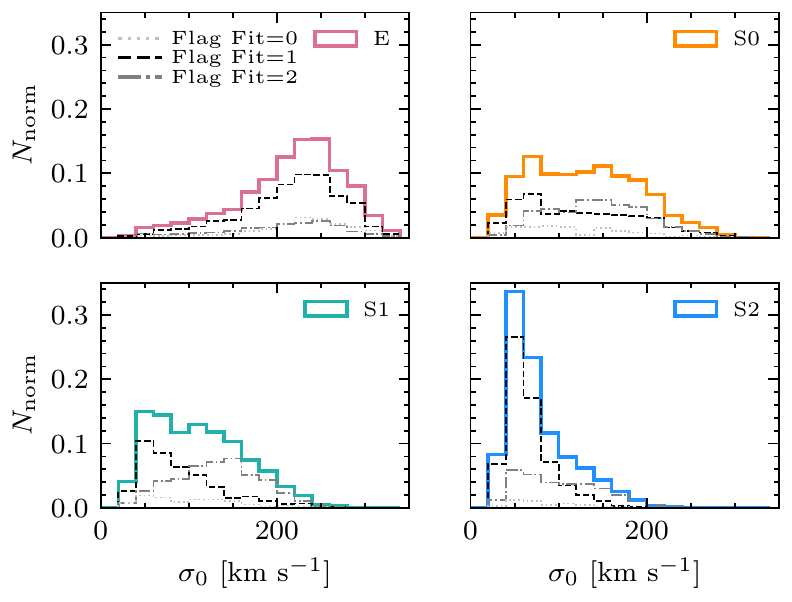}\par 
    \caption{Same as previous figure, but for central velocity dispersion $\sigma_0$.  Es tend to have large $\sigma_0$, whereas for S2s $\sigma_0$ tends to be very small. S1  with preferred 1-component fit (FLAG$\_$FIT=1) have smaller $\sigma_0$.}
    \label{fig:Sig0_flag_fit}
\end{figure}

\begin{figure}
    \includegraphics[width=0.9\linewidth]{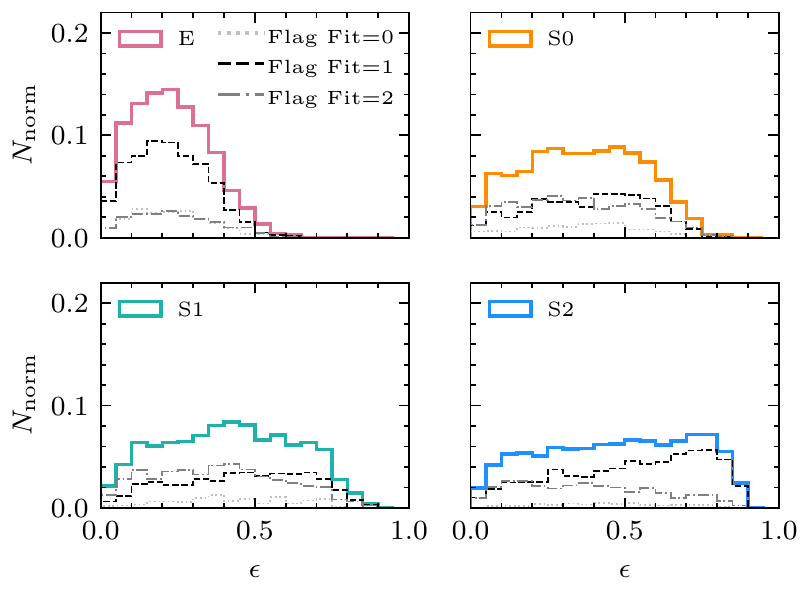}\par 
    \caption{Same as previous figure, but for observed $\epsilon\equiv 1 - b/a$.  Es tend to be round, whereas S2s have a wide range of $\epsilon$ as expected for inclined disks.  This correlation with morphology is striking because PyMorph b/a played no role in the classification.}
    \label{fig:epsilon_flag_fit}
\end{figure}

\section{Combining the two catalogues}\label{sec:photmorph}

In this section, we consider the benefits of combining the MPP-VAC-DR17 with the MDLM-VAC-DR17.  Table~\ref{tab:fit} shows the frequency of  FLAG$\_$FIT  for galaxies separated by morphological classes.  E and S2 galaxies tend to be better described by 1-component fit, while S0 and S1 show a mixture of 1 and 2-components. This does not depend on whether we use T-Type or visual classifications to define the morphological class. On the other hand, more than half of the galaxies classified as ETGs or LTGs are better described by 1-component fit (this is due to the ETGs being a mix of E and S0 and LTGs a mix of S1 and S2).

Figures~\ref{fig:nBT_flag_fit}--\ref{fig:epsilon_flag_fit} show how the distributions of $n$, B/T, luminosity, central velocity dispersion and $\epsilon\equiv 1 - b/a$ depend on morphology and FLAG$\_$FIT. The morphological classes shown in the following are based on the VC values, with an additional separation between S1 and S2 at  T-Type=3.

The figures show that Es and S2s tend to be dominated by galaxies better described by 1-component (FLAG$\_$FIT=1), with a S{\'e}rsic index peaking around $\sim 4-6$ and $\sim 1$, respectively. The S2s with FLAG$\_$FIT=1 also tend to have larger $\epsilon$.

The S0s and S1s tend to have more similar numbers of galaxies better described by 1 or 2 components (FLAG$\_$FIT=1 or 2), with 1-component objects tending to be less luminous and to have smaller $\sigma_0$. 
The B/T distributions of galaxies better described by 2-components (FLAG$\_$FIT=2) also show the expected trends: as one goes to later types, the peak of the distribution (and its skewness) shifts towards lower B/T.  
Since the PyMorph fits played no role in the morphological classification, the correspondence between FLAG$\_$FIT and morphology in these figures is remarkable, and is why we believe FLAG$\_$FIT should be used in scientific analyses of our photometric catalog.

\section{Comparison with Galaxy Zoo}\label{sec:zoo}
We now compare our MDLM Deep Learning morphologies with those of the GZ2 provided by \cite{Willett2013}, in the same format as Figures~24-26 in F19.  We use the "weighted fraction" GZ2 probability P$_{\rm Smooth}$ which is sometimes used as a proxy for "early-type" (ETGs) and "late-type" (LTGs) galaxies. 

\begin{figure}
    \includegraphics[width=0.9\linewidth]{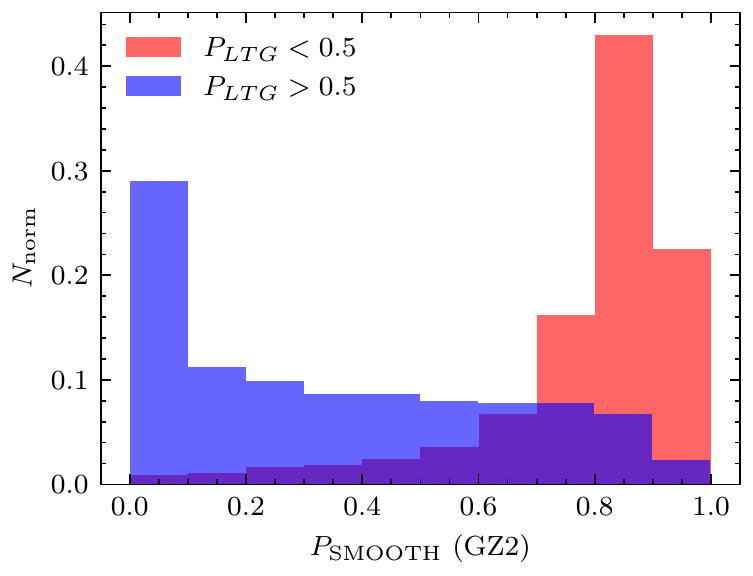}
    \includegraphics[width=0.9\linewidth]{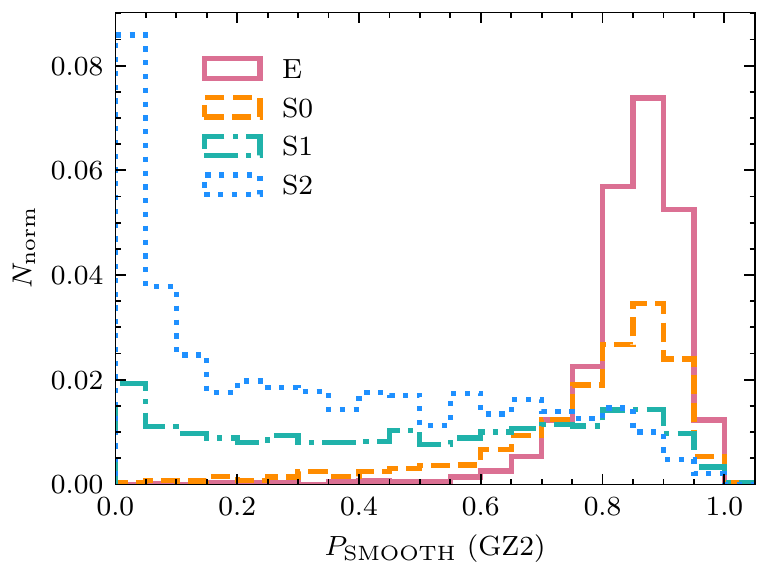}
    \caption{Distribution of P$_{\rm smooth}$ from GZ2 for DR17 galaxies divided according to our P$_{\rm LTG}$ (top) and T-Type + P$_{\rm S0}$ (bottom).  Most galaxies with P$_{\rm smooth} \ge 0.6$ have P$_{\rm LTG}\le 0.5$ and tend to be Es or S0s, although there is a non-negligible fraction of Ss.  
    }
    \label{fig:GZ2}
\end{figure}

The top panel of Figure~\ref{fig:GZ2} shows that objects with P$_{\rm LTG}\le 0.5$ -- i.e. that are unlikely to be LTGs -- tend to have large P$_{\rm smooth}\ge 0.6$ (their images are smooth, with no disk features).  Although we do not show it here, objects with P$_{\rm LTG}\le 0.5$ tend to have P$_{\rm disk}\le 0.3$ (i.e., they are unlikely to be disks), as expected.  The bottom panel shows that, although most galaxies with P$_{\rm smooth}\ge 0.6$ are dominated by Es or S0s, there is a significant fraction ($\sim$30$\%$) of objects which are Ss .

\begin{figure}
    \includegraphics[width=0.9\linewidth]{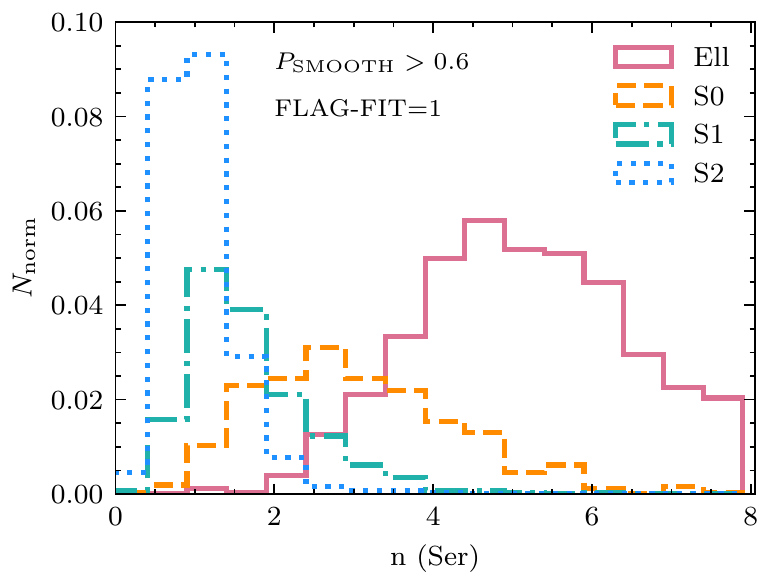}
    \includegraphics[width=0.9\linewidth]{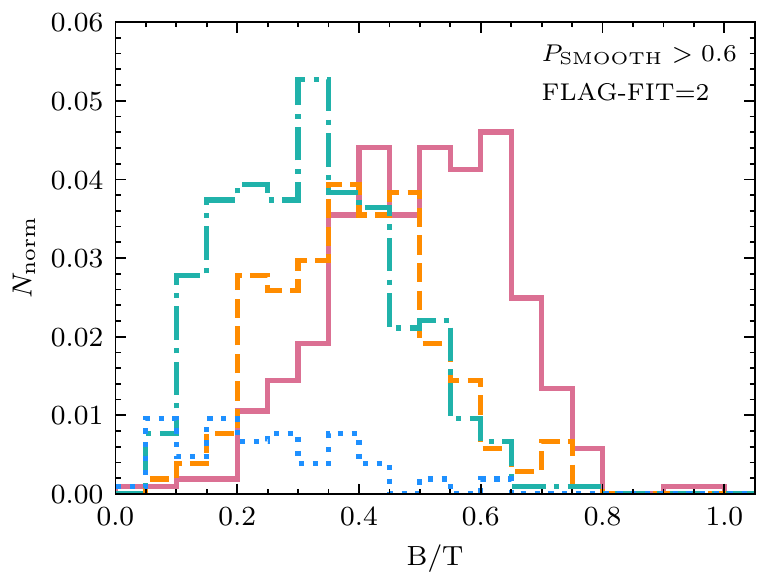} 
    \caption{Distribution of $r$-band S{\'e}rsic index $n$  for galaxies better described by 1-component (FLAG$\_$FIT=1, top) and B/T  for galaxies better described by 2-components (FLAG$\_$FIT=2, bottom), for galaxies with GZ2 P$_{\rm smooth} > 0.6$.  Objects with small $n$ or B/T tend to be S, confirming that our morphological classification is correct.  Results for  P$_{\rm disk}\le 0.3$ from GZ2 are nearly identical, so we have not shown them here.}
\label{fig:GZ2smooth}
\end{figure}

To check if our S1 and S2 classifications at P$_{\rm smooth}\ge 0.6$ are incorrect, Figure~\ref{fig:GZ2smooth} shows the distribution of $n$ for 1-component galaxies (FLAG$\_$FIT=1) and B/T for 2-component galaxies (FLAG$\_$FIT=2).  The Es clearly have larger $n$ and B/T, and the Ss clearly have $n\sim 1$ and lower B/T (neither $n$ nor B/T played a role in determining T-Type or P$_{\rm Smooth}$).  This strongly suggests that our classifications are appropriate, so
(a) the presence of Ss with P$_{\rm Smooth} > 0.6$ implies that conclusions about Es that are based on GZ2 P$_{\rm Smooth}$ should be treated with caution;
(b) selecting Es based on our MDLM T-Type classifications is much more robust than selecting on GZ2 P$_{\rm Smooth}$.  

In their analysis of DR15, F19 showed that selecting objects with P$_{\rm disk}<0.3$ produces almost identical results as Figure \ref{fig:GZ2smooth}.  This remains true in DR17, so we have not shown it explicitly.

\section{Summary and Conclusions}

We have presented the  MaNGA PyMorph photometric VAC (MPP-VAC-DR17) and the MaNGA Deep Learning Morphological Value Added Catalogue (MDLM-VAC-DR17) for the final data release of the MaNGA survey (which is part of the SDSS Data release 17 – DR17). 

The MPP-VAC-DR17 is an extension of the MPP-VAC-DR15 to include all the galaxies in the final MaNGA release. It provides photometric parameters for 2D surface brightness profiles for 10293 observations (of which 10127 are unique galaxies) in the $g, r$ and $i$ bands. The MPP-VAC  is identical to the one presented in F19 and its content is detailed in Table 1 of such paper. The only difference with the MPP-VAC-DR15 is the definition of the position angle (PA), given in this catalogue with respect to the camera columns in the SDSS “fpC” images. The 2D light profile fittings are derived both for Sersic and SerExp models. The catalogue contains a flagging system that indicates which fit is to be preferred for scientific analyses  (FLAG$\_$FIT=1 for Sersic, FLAG$\_$FIT=2 for SerExp, FLAG$\_$FIT=0 when both are acceptable). We urge users to pay attention to the preferences expressed by this flag since some fits may be unreliable.

The MDLM-VAC-DR17 is also an extension of the MDLM-VAC-DR15 presented in F19 and includes exactly the same entries as MPP-VAC-DR17. The MDLM-VAC-DR17 implements some changes  compared to the previous release and its content its detailed in Table \ref{tab:morph}. The main improvements of the new release are:
\begin{itemize}
\item [(i)] the low-end of the T-Types is better recovered, with a smaller bias $b$ compared to the previous version, thanks to a change in the CNN architecture (an additional dense layer was added, see section \ref{sect:training}).
\item [(ii)] a new binary model P$_{\rm LTG}$, which separates ETGs from LTGs in a complementary way to the T-Type, especially at intermediate T-Types where the scatter is larger.
\item [(iii)] all the classifications are trained using $k$-folding (with $k=15$ for the T-Type and $k=10$ for the rest of the models) and the value reported in the catalogue is the average of the $k$ models.
\item [(iv)] we report the standard deviation of the outputs of the $k$ models, which can be used as a proxy for their uncertainties (see Figures \ref{fig:Sigma-P_LTG}, \ref{fig:ROC-edge}).
\item  [(v)] a visual classification (VC=1 for E, VC=2 for S0, VC=3 for S/Irr) and visual flag (VF=0 for reliable classification, VF=1 for uncertain classifications) is also included.
\end{itemize}

By combining the different classification models we find:

\begin{itemize}
    \item Galaxies having inconsistent T-Type and P$_{\rm LTG}$ classifications  tend to be faint  and small galaxies with  small central velocity dispersion, i.e., they are difficult to classify. In general, they share some properties with LTGs but they have no obvious spiral features.
    
    \item The larger discrepancy between the visual classification and the one provided by the combination of the T-Type and the P$_{\rm S0}$ (as defined in section \ref{sect:P-LTG}) is for galaxies classified as S0 by the former and S by the latter (see Table \ref{tab:VC}). This fraction is reduced from 10\% to 1\% when only galaxies with VF=0 are considered.
        \item The larger discrepancy between the visual classification and  P$_{\rm LTG}$ is for galaxies classified as S by the former and ETG by the  latter (see Table \ref{tab:VC}). This fraction is reduced from 16\% to 5\% when only galaxies with VF=0 are considered.
\end{itemize}

By combining the two catalogues MPP-VAC-DR17 and  MDLM-VAC-DR17 and despite the changes to the morphological classification,  we find similar results as in F19. Namely:

\begin{itemize}
    \item There is a strong correlation between the morphological classification and the values of $n$, $n_{\rm bulge}$ and B/T (see figure \ref{fig:nBT_flag_fit}).
    \item E galaxies tend to be bright (more negative M$_r$ values), have large central velocity dispersion $\sigma_0$ and small ellipticity $\epsilon$ while the trend is the opposite for the S galaxies, especially S2 (see Figures \ref{fig:Mr_flag_fit}--\ref{fig:epsilon_flag_fit}). 
     \item Separating galaxies according to the FLAG$\_$FIT we observe that Es and S2s tend to be dominated by galaxies better described by 1-component fit (FLAG$\_$FIT=1). On the other hand, the S0s and S1s tend to have more similar numbers of objects described by 1 or 2-component fit (FLAG$\_$FIT=1 and 2), with 1-component objects tending to be less luminous and having smaller $\sigma_0$. 
     \item Since the PyMorph fits played no role in the morphological classification, the correspondence between FLAG$\_$FIT and morphology is remarkable:  FLAG$\_$FIT should be used in scientific analyses of our photometric catalog.
     \item We find a significant fraction of S galaxies with P$_{\rm Smooth} > 0.6$ from GZ2 (Figure \ref{fig:GZ2}). Most of these galaxies have  have $n\sim 1$ and low B/T (Figure \ref{fig:GZ2smooth}), consistent with being disk galaxies.  Therefore, as F19 noted previously, conclusions about Es that are based on GZ2 P$_{\rm Smooth}$ should be treated with caution.

\end{itemize}

\section*{Acknowledgements}

The authors thank the referee for useful comments which helped to improve the quality of the paper. This work was supported in part by NSF AST-1816330. 
HDS acknowledges support from PIE2018-50E099 project: Cross-field research in space sciences.
The authors gratefully acknowledge the computer resources at Artemisa, funded by the European Union ERDF and Comunitat Valenciana as well as the technical support provided by the Instituto de Física Corpuscular, IFIC (CSIC-UV).

Funding for the Sloan Digital Sky 
Survey IV has been provided by the 
Alfred P. Sloan Foundation, the U.S. 
Department of Energy Office of 
Science, and the Participating 
Institutions. SDSS-IV acknowledges support and 
resources from the Center for High 
Performance Computing  at the 
University of Utah. The SDSS 
website is www.sdss.org.

SDSS-IV is managed by the 
Astrophysical Research Consortium 
for the Participating Institutions 
of the SDSS Collaboration including 
the Brazilian Participation Group, 
the Carnegie Institution for Science, 
Carnegie Mellon University, Center for 
Astrophysics | Harvard \& 
Smithsonian, the Chilean Participation 
Group, the French Participation Group, 
Instituto de Astrof\'isica de 
Canarias, The Johns Hopkins 
University, Kavli Institute for the 
Physics and Mathematics of the 
Universe (IPMU) / University of 
Tokyo, the Korean Participation Group, 
Lawrence Berkeley National Laboratory, 
Leibniz Institut f\"ur Astrophysik 
Potsdam (AIP),  Max-Planck-Institut 
f\"ur Astronomie (MPIA Heidelberg), 
Max-Planck-Institut f\"ur 
Astrophysik (MPA Garching), 
Max-Planck-Institut f\"ur 
Extraterrestrische Physik (MPE), 
National Astronomical Observatories of 
China, New Mexico State University, 
New York University, University of 
Notre Dame, Observat\'ario 
Nacional / MCTI, The Ohio State 
University, Pennsylvania State 
University, Shanghai 
Astronomical Observatory, United 
Kingdom Participation Group, 
Universidad Nacional Aut\'onoma 
de M\'exico, University of Arizona, 
University of Colorado Boulder, 
University of Oxford, University of 
Portsmouth, University of Utah, 
University of Virginia, University 
of Washington, University of 
Wisconsin, Vanderbilt University, 
and Yale University.

\section*{Data Availability}

The catalogues described in this article are part of the final data release of the MaNGA survey and will be released as part of the SDSS DR17. The catalogues are available in https:$//$www.sdss.org$/$dr17$/$data$\_$access$/$value-added-catalogs. The code used for the deep learning algorithm may be shared upon request.





\bibliographystyle{mnras}
\bibliography{sample} 





\bsp	
\label{lastpage}
\end{document}